\documentclass[conference]{IEEEtran}
\usepackage{amsmath,amssymb,amsfonts}
\usepackage{algorithmic}
\usepackage{textcomp}
\usepackage{xcolor}
\def\BibTeX{{\rm B\kern-.05em{\sc i\kern-.025em b}\kern-.08em
    T\kern-.1667em\lower.7ex\hbox{E}\kern-.125emX}}
\usepackage{graphicx}

\usepackage[export]{adjustbox}
\usepackage{booktabs}
\usepackage{cite}

\ifCLASSINFOpdf
\else
\fi

\usepackage{amsmath}
\begin{document}
%
\title{Pushing the Accuracy Limit of Foundation Neural Network Models with Quantum Monte Carlo Forces and Path Integrals}

\author{\IEEEauthorblockN{
Anouar Benali\IEEEauthorrefmark{1},
Thomas Plé\IEEEauthorrefmark{2},
Olivier Adjoua \IEEEauthorrefmark{2},
Valay Agarawal \IEEEauthorrefmark{3},
Thomas Applencourt \IEEEauthorrefmark{4},
Marharyta Blazhynska\IEEEauthorrefmark{2},
\\
Raymond Clay III \IEEEauthorrefmark{5},
Kevin Gasperich \IEEEauthorrefmark{1},
Khalid Hossain \IEEEauthorrefmark{4},
Jeongnim Kim \IEEEauthorrefmark{6},
Christopher Knight \IEEEauthorrefmark{7},
Jaron T. Krogel \IEEEauthorrefmark{8},\\
Yvon Maday \IEEEauthorrefmark{9},
Maxime Maria \IEEEauthorrefmark{10},
Matthieu Montes \IEEEauthorrefmark{11},
Ye Luo \IEEEauthorrefmark{7},
Evgeny Posenitskiy \IEEEauthorrefmark{12},
Corentin Villot \IEEEauthorrefmark{12},\\
Venkatram Vishwanath\IEEEauthorrefmark{4},
Louis Lagardère \IEEEauthorrefmark{2,12}\IEEEauthorrefmark{12},
Jean-Philip Piquemal\IEEEauthorrefmark{2}\IEEEauthorrefmark{12}
}
\IEEEauthorblockA{
    \IEEEauthorrefmark{1}Qubit Pharmaceuticals Inc, Boston, MA 02116, USA}
    \IEEEauthorblockA{\IEEEauthorrefmark{2} Laboratoire de Chimie Théorique, Sorbonne Université, UMR 7616 CNRS,75005 Paris, France\\}
    \IEEEauthorblockA{\IEEEauthorrefmark{3} Department of Chemistry, University of Chicago, Chicago, IL 60637, USA}
    \IEEEauthorblockA{\IEEEauthorrefmark{4} Argonne Leadership Computing Facility, Argonne National Laboratory, Argonne, IL 60439, USA}
    \IEEEauthorblockA{\IEEEauthorrefmark{5}
    HEDP Theory Department, Sandia National Laboratories,Albuquerque, NM 87185, USA}
    \IEEEauthorblockA{\IEEEauthorrefmark{6}
    Intel Corporation, Hillsboro, OR 987124, USA}
    \IEEEauthorblockA{\IEEEauthorrefmark{7}Computational Science Division, Argonne National Laboratory, Argonne, IL 60439, USA}
    \IEEEauthorblockA{\IEEEauthorrefmark{8}Materials Science and Technology Division, Oak Ridge National Laboratory, Oak Ridge, TN 37831, USA}
    \IEEEauthorblockA{\IEEEauthorrefmark{9}Sorbonne Université, Université Paris Cité, CNRS, INRIA, LJLL UMR 7598, Paris 75005, France}
\IEEEauthorblockA{\IEEEauthorrefmark{10}Université de Limoges, XLIM UMR 7252 CNRS, 87060 Limoges, France}
\IEEEauthorblockA{\IEEEauthorrefmark{11} Conservatoire National des Arts et Métiers et Institut Universitaire de France, Laboratoire GBCM, 75003 Paris, France}
\IEEEauthorblockA{\IEEEauthorrefmark{12}Qubit Pharmaceuticals,75014 Paris, France\\Corresponding author: \textcolor{blue}{jean-philip.piquemal@sorbonne-universite.fr}}
    
 }


%


\maketitle

\begin{abstract}

We propose an end-to-end integrated strategy to produce highly accurate quantum chemistry (QC) synthetic datasets (energies and forces) aimed at deriving Foundation Machine Learning models for molecular simulation. Starting from Density Functional Theory (DFT), a "Jacob's Ladder" approach leverages computationally-optimized layers of massively GPU-accelerated software with increasing accuracy. Thanks to Exascale, this is the first time that the computationally intensive calculation of Quantum Monte Carlo forces (QMC), and the combination of multi-determinant QMC energies and forces with selected-Configuration Interaction wavefunctions, are computed at such scale at the complete basis-set limit. To bridge the gap between accurate QC and condensed-phase Molecular Dynamics, we leverage transfer learning to improve the DFT-based FeNNix-Bio1 foundation model. The resulting approach is coupled to path integrals adaptive sampling quantum dynamics to perform nanosecond reactive simulations at unprecedented accuracy. These results demonstrate the promise of Exascale to deepen our understanding of the inner machinery of complex biosystems.
\end{abstract}


%
\IEEEpeerreviewmaketitle

\section{Justification for Prize}
\begin{itemize}
    \item Record Diffusion Quantum Monte Carlo (DMC): energies-only, energies/forces and Multideterminant DMC: energies/forces; double-precision performance: 124, 62, 52 PFLOPS; 60\%, 30\%, 25\% Aurora's theoretical peak; 12,288 GPUs. 
\item Record time-to-solution for full quantum dynamics foundation machine learning model simulation. Time-to-solution: 0.44ns/day mixed-precision (1024 GPUs, Jean Zay; 5760 GPUs Aurora).

\end{itemize}

\section{Performance Attributes}

{\scriptsize
\begin{center}
\begin{tabular}{l l} 

 \toprule
 Performance attribute & Our submission  \\
 \midrule
 Category of achievement  & Time-to-solution, scalability \\ 
 Type of methods used   &  Quantum Chemistry, ML potentials MD \\
   Results reported on basis of & Whole application including I/O \\
 Precision reported & Double precision, mixed precision \\
 System scale & Measured on full system \\
 Measurements  & Timers, FLOP count\\
\bottomrule
\end{tabular}
\end{center}
}

\section{Overview of the problem}
Simulating the structural dynamics of biological systems at full quantum accuracy is a daunting task requiring the precise modeling of the Born-Oppenheimer (BO) electronic potential energy surface\cite{Helgaker20082008} while including nuclear quantum effects (NQEs) in the dynamics to accurately reproduce condensed phase properties\cite{markland2018nuclear}. Up to now, achieving such high-fidelity simulations has been prevented by the computational cost of first-principles quantum approaches that can only reach maximum accuracy on the BO surface of small few-atom systems\cite{Helgaker20082008}. Full quantum accuracy MD remains therefore a Graal for the molecular simulation community. In this context, neural networks potentials\cite{yang2024machine} have started to gain speed and precision offering a clear alternative to ab initio molecular dynamics (AIMD)\cite{marx2009ab} and recently even to force fields simulations\cite{MACKERELL200591} thanks to GPU-accelerated high performance supercomputing\cite{jia2020pushing}. Thus, the field evolves very fast, and larger general-purpose foundation neural network models dedicated to atomistic Molecular Simulation have been recently introduced\cite{batatia2024foundationmodelatomisticmaterials}. However, these ML approaches cannot perform computations at an accuracy better than their reference data. Consequently, the choice of the synthetic quantum chemistry approaches used for the reference datasets and their capability to approximate the Schrödinger equation have a considerable impact. Additionally, once reaching high precision on the BO energies, it becomes paramount to accurately reproduce experimental observations. To take full advantage of the high accuracy achieved by neural networks, one must take into account for NQEs explicitly\cite{markland2018nuclear,mauger2021nuclear} via some form of quantum dynamics. This is mandatory for the accurate prediction of condensed-phase properties. Overall, the rise of exascale computing offers the possibility of designing a new, multi-level pipeline for the generation of accurate quantum chemistry datasets. This will enable to derive ML models capable of performing large scale quantum dynamics molecular simulations towards accurately predicting the structural dynamics and chemical reactivity of highly complex biosystems.
\section{Current State of the art}
\subsubsection{Quantum Chemistry Datasets}
 The present consensus for the computation of reference quantum chemistry datasets is to use DFT potential energy surfaces which provide a reasonable accuracy at an affordable cost. However DFT's precision strongly depends on the choice of the functional, which is by essence non-exact and therefore not fully transferable across the chemical space. Therefore, there is a need for more exact ab initio methods. 
 For several decades CCSD(T)\cite{RAGHAVACHARI1989479,schafer2024} often termed the "gold standard", has been widely employed. Nonetheless, CCSD(T) exhibits steep $O(N^7)$ scaling, and its local approximations (e.g., DLPNO-CCSD(T)) rely on domain truncations that are typically benchmarked against known data but can fail if the chosen domains omit critical electron correlation\cite{schafer2024}. As a result, such local-correlation variants, while computationally attractive for large systems, can yield unpredictable errors outside the familiar chemical space. Similarly, Density Matrix Renormalization Group (DMRG) offers high accuracy for strongly correlated systems but remains challenging for broad chemical data generation due to the need for carefully chosen active spaces.\cite{legeza2025}\\
By contrast, Quantum Monte Carlo (QMC) methods\cite{Reynolds_1982}, and particularly Diffusion Monte Carlo (DMC)\cite{Reynolds1990,foulkes01}, have proven capable of near-exact ground-state energies across wide chemical spaces\cite{foyevtsova14,shin17,shin2017,kylanpaa17,shin18,Morales2012,Benali_2014,benali16,Kolorenvc_2011,Dubecky_2016}. Large-scale QMC studies focused on benchmark energies\cite{Benali2020,khan2024}, but until recently lacked an efficient pathway to obtain forces, rendering them less suitable for force field generation. Computing forces in Diffusion Monte Carlo (DMC) remains inherently challenging due to the fixed-node approximation and the intrinsic stochastic nature of the algorithm. Fortunately, over the past several years, a variety of technical\cite{Filippi2000}, mathematical\cite{Wagner2014}, and performance improvements\cite{Filippi2016} to the original Zero-Variance Zero-Bias force estimator framework of Assaraf and Caffarel\cite{Caffarelforces2000} have made QMC forces usable and trustworthy for large-scale applications.\\

Unlike traditional quantum chemistry methods, which rely on finite basis sets and must extrapolate to the complete basis set (CBS) limit or depend on error cancellation and ad hoc corrections, QMC,  and particularly DMC, operates directly in real space. As a result, it is inherently free from basis set incompleteness error and provides results effectively in the CBS limit without the need for basis-set tricks or extrapolation schemes.

Generating multideterminant expansions to reduce nodal-surface errors has proven to be challenging. Methods such as CASSCF have been recognized for decades as effective but require extensive prior knowledge of the target system’s chemical properties, which limits their suitability for broad machine-learning applications. More recently, Caffarel and co-workers\cite{Caffarel2013} revived the concept of selected Configuration Interaction (sCI)\cite{HURON1974277}, automating the construction of multideterminant expansions in a largely black-box manner and enabling various sCI flavors\cite{Tubman2016,Holmes_2017}. As a multideterminant trial wavefunction, sCI systematically improves nodal accuracy. Coupled with an optimized Jastrow factor, it resolves both static and dynamic correlations by addressing strong-correlation effects through the sCI expansion while capturing dynamical correlations in the Jastrow\cite{Benali2020,Scemama_2018b,Loos_2018b,Loos_2020b,Loos_2020a,Scemama_2019,Loos_2019,Neuscamman2020,neuscamman2019,Garner2020}. In turn, QMC forces can validate or directly compute sCI forces, establishing a synergistic approach with high reliability\cite{Slootman2024}.

Nevertheless, the computational cost of these methods remains substantial, restricting routine usage for many years. Over the past two decades, the community has extensively developed QMCPACK\cite{QMCPACK1,QMCPACK2,QMCPACK3,QMCPACK4} to exploit leadership-class supercomputers—from IBM Blue Gene architectures to exascale systems such as Aurora and Frontier at Argonne and Oak Ridge National Laboratories respectively. Early efforts prioritized a GPU-based energy estimator, particularly via efficient B-spline orbital evaluations on massively parallel platforms, targeting material sciences application. Building on this foundation, the present work introduces new optimizations and data layouts that optimize the LCAO branch of QMCPACK for GPU acceleration for molecular applications.\\
In parallel, a modern sCI program has been developed with GPU-friendly data structures and offloaded core operations in preparation for large-scale parallelism. Taken together, these advances enable QMC–sCI workflows on exascale platforms, providing the high-fidelity datasets needed for force-field generation and facilitating the study of previously intractable systems in chemical and materials research.
 \subsection{\textbf{Foundation Machine Learning Models for atomistic molecular dynamics Simulations}}
 The application of Neural Networks to MD is recent and started with the availability of suitable neural architectures ~\cite{behler2007generalized, bartok2010gaussian}. Since then, numerous neural network potentials (NNPs) have been proposed \cite{chmiela2018towards,bigi2023wigner,shakouri2017accurate,smith2017ani,schutt2017quantum,schutt2017schnet,gilmer2017neural,lubbers2018hierarchical,zubatyuk2019accurate, zaverkin2020gaussian,qiao2021unite,unke2021spookynet,gasteiger2021gemnet,batzner20223,musaelian2022learning,grisafi2019incorporating,grisafi2021multi,ANI2X, AIMNET, unke2024biomolecular,pinheiro2021choosing,kocer2022neural,yang2024machine}. As NNPs' high performance implementations reached a mature stage \cite{lammpsnn,jia2020pushing,inizan2022scalable}, specialized libraries ~\cite{schutt2023schnetpack,dral2024mlatom, zeng2023deepmd, ple2024fennol} are now available to developers facilitating the design of new NNPs. Very recently, following the advances of the AlphaFold \cite{alphafold} and RoseTTAFold \cite{rosettafold} protein structure foundation models, a shift in the developments started with the introduction of the MACE-MP-0 foundation model for MD. \cite{batatia2024foundationmodelatomisticmaterials}. MACE is a single general-purpose ML potential, designed for material science atomistic simulations. Its capabilities range from single molecule to full materials simulations, and enables modeling of chemical reactivity. However, since MACE and its associated neural architecture are primarily designed for materials, its applicability to condensed phase simulations of large complex biological molecules remains limited both in term of accuracy and computational performances. For these reasons, some of us recently introduced the FeNNix-Bio1 Foundation Model\cite{FennixBio} for atomistic MD simulations going beyond the initial FeNNix equivariant NNP\cite{D3SC02581K}. Thus, FeNNIx-Bio1 is able to perform reactive atomistic condensed phase MD simulations of biological systems addressing simultaneously the issues of the computational requirements generated by the biological timescales and the difficult transferability of charged-species. Finally, it is also able to compute meaningful condensed-phase properties (free energies, solvent properties,...) as it takes explicitly into account NQEs \cite{ceriotti2016nuclear,mauger2021nuclear,ple2023routine}. 
 \section{Innovations}
\subsection{Dataset Generation: High Performance Quantum Chemistry Pipeline}
\begin{figure}
        \adjustimage{width=0.45\textwidth,center}{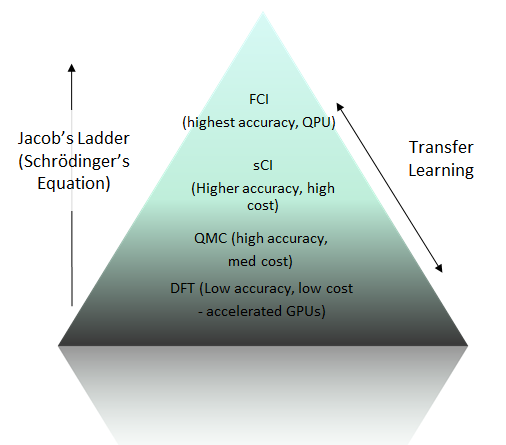}
    \caption{Jacob's Ladder for quantum chemistry methods towards the full Schrödinger’s Equation. Acronyms: FCI= Full Configuration Interaction, sCI= Selected Configuration Interaction; QMC= Quantum Monte-Carlo; DFT=Density Functional Theory; QPU= Quantum Processor Unit.}
    \label{fig:jacob}
\end{figure}

We first introduce our GPU-accelerated quantum chemistry database generation Exascale protocol which has been optimized to able to produce the required energies and forces. Organized as a ``Jacob's Ladder'' approach (see Figure ~\ref{fig:jacob}) with methods of increasing complexity, it leverages 3 different layers of computationally-optimized massively parallel high performance softwares dealing with increasing levels of accuracy. Note that exact, i.e. Full-Configuration interaction (or Full-CI) computations in the CBS limit, numerically reaching the quality of the electronic time-independent, non-relativistic, Schrödinger’s equation quality computations are mostly out of reach of classical computers due to the existence of a complexity Exponential Wall problem\cite{RevModPhys.71.1253,FCIlimit}. The multi-level GPU-accelerated computational strategy proposed here aims at reaching the chemical accuracy which is the required level of precision needed to perform realistic chemical, i.e. comparable to experiment, predictions. It implies errors below 1 kcal/mol (1.59 mH/particle) with respect to the exact Schrödinger total energy. Since forces are the cornerstone of the design of ML potentials, each proposed level includes the efficient evaluation of the energy gradients.   

\subsubsection{The SPICE2(+)-ccECP dataset}
Testing the computational pipeline, we computed a first training set denoted SPICE2(+)-ccECP.
It used the molecular configurations of the SPICE2 database\cite{Spice2_1,eastman2023spice}, which we extended with a carefully selected subset of 100k configurations from the ANI-2X dataset\cite{ANI2X}. These latter were selected via a round of active learning using the uncertainty given by a preliminary version of the FeNNix-Bio1 model and recomputed using our selected level of theory. This combined dataset provided comprehensive coverage of the conformational space relevant to biomolecular systems, including diverse functional groups, a wide range of environmental conditions, and various non-equilibrium configurations critical for modeling reactive processes.
Diverging from the original SPICE2, we combined in the present DFT work with correlation-consistent effective core potentials (ccECP)\cite{ccECP} and their companion aug-cc-pVTZ basis sets\cite{ccECP}, as implemented in the PySCF package\cite{PySCF}. This choice conserves accuracy while increasing the computational efficiency compared to the explicit inclusion of core electrons while enableing us to extend the SPICE2 dataset to post Hartree-Fock methods such as QMC where the use of valence-only calculations is crucial. The dataset encompasses 2,108,628 conformations that were all treated (energies and forces) at the DFT level. Among these latter, 20,251 conformations were computed using QMC (energies and forces), 20,338 QMC (energies-only) and 2000 (energies and forces) with multideterminant QMC with selected CI wavefunctions.
\subsubsection{\textbf{Density Functional Theory}}
\begin{itemize}
    \item \textbf{Summary of Contributions}
\end{itemize}
For the first layer of our quantum chemistry database generation protocol, we performed DFT calculations using the PySCF code\cite{pyscf_cpu,Pyscf_gpu1,Pyscf_gpu2}, specifically leveraging the GPU-accelerated implementation. We strategically allocated computational resources based on system size. For molecules up to 175 electrons (approximately 1.3 million configurations), calculations were executed on the Leonardo supercomputer at Cineca (Italy), larger molecular systems with up to 275 electrons (approximately 0.8 million configurations) were computed on the Polaris supercomputer at Argonne National Laboratory. Both supercomputers feature NVIDIA A100 GPUs-40Gb. The largest molecules with more than 275 electrons (65,521 conformations) were computed on Aurora Supercomputer which features Intel Xe GPUs. A significant challenge for large-scale DFT calculations is the computation and storage of electron repulsion integrals (ERIs), which cannot be generated on-the-fly on GPUs for larger molecules due to memory limitations of the A100-40Gb cards. The conventional CPU implementation would require over 6 hours per molecule on AMD Epyc 7303 processors and up to 300GB of temporary storage per calculation, making the complete dataset generation computationally infeasible. To overcome these limitations, we employed our gpu4mrh implementation, which optimizes memory usage and enables efficient offloading of integral calculations to GPUs even for large molecular systems.

We employed the $\omega$B97M-D3BJ functional\cite{Mardirossian2016,Mardirossian2017,Grimme2011}, which provides an excellent balance between accuracy and computational efficiency for organic and biological systems. This range-separated hybrid meta-GGA functional incorporates D3BJ dispersion corrections to accurately capture non-covalent interactions critical for biological systems. All calculations utilized correlation-consistent effective core potentials (ccECP)\cite{Bennett2017,Annaberdiyev2018} to reduce computational cost while maintaining high accuracy for valence electrons.

\begin{itemize}
    \item \textbf{Algorithmic Innovations}
\end{itemize}
To optimize computational resources while ensuring sufficient accuracy, we employed an adaptive quadrature grid strategy. For energy calculations, we utilized a grid level of 3, while for gradient calculations (forces), we increased the grid density to level 4 to ensure accurate force evaluation. This strategy significantly reduced computational cost and memory usage while maintaining chemical accuracy in both energies and forces.\\
The DFT calculations were distributed across the Nvidia A100 nodes systems (Polaris/Leonardo) with one molecule assigned per GPU, using 4 GPUs per node. This distribution strategy optimized the trade-off between parallel efficiency and memory requirements. Specifically, the DFT calculations achieved approximately 16 TFLOPS per node (40\% of theoretical peak) for smaller molecules (0–100 electrons), 20 TFLOPS per node (50\% of theoretical peak) for intermediate-sized molecules (100–200 electrons), and 28 TFLOPS per node (70\% of theoretical peak) for larger molecules (200–250+ electrons). Each DFT simulation was distributed across 128 nodes, with 4 Nvidia A100 GPUs per node (512 GPUs per simulation in total). This large-scale deployment was essential, as it was the only feasible strategy to generate the extensive dataset of 2.1 million molecular configurations within the allocated time frame. Specifically, the simulations achieved peak performances of approximately:
\begin{itemize}
    \item 1.98 PFLOPS) for smaller molecules (0–100 electrons),
    \item 2.48 PFLOPS for intermediate-sized molecules (100–200 electrons),
    \item 3.48 PFLOPS for larger molecules (200–250+ electrons).
\end{itemize}
These performances represent approximately 40–70\% of the peak theoretical capability of the utilized hardware nodes.\\
On Aurora, we used PySCF coupled with the gpu4mrh code that leverages the Intel Xe architecture through efficient data layout patterns and kernel optimizations to maximize use of matrix-multiply operations central to quantum chemistry calculations. The gpu4mrh code, initially developed to accelerate multi-configurational electronic structure methods, uses a lightweight Python/C++ abstraction layer to seamlessly offload fundamental operations from PySCF (namely, formation of the Coulomb and exchange matrices) to GPUs via CUDA, HIP, or SYCL and respective vendor-optimized math libraries (e.g. MKL on Aurora). With an efficient hashing strategy to minimize data-transfer GPU-host data transfers, the majority of the GPU runtime is spent computing single and batched matrix-matrix products concurrently across multiple GPUs in a compute node. The gpu4mrh code was used as a drop-in replacement enabling SCF energy calculations to leverage the full 768 GB of HBM from the 6 GPUs on each Aurora node. 
The implementation cost is dominated by several dense gemm operations, so it is able to make use of vendor-optimized BLAS routines and achieves high GPU utilization for the systems considered here.

In total, we performed computed 2,1 million molecular configurations, requiring approximately 20k nodes-hours on Leonardo, 35k node-hours on Polaris and 25k nodes hours on Aurora. The resulting database of energies and forces serves is the foundation layer for our multi-level quantum chemistry approach.\\
In addition to providing the first level of our Jacob's Ladder approach, these DFT calculations served as essential inputs for our higher-level quantum chemistry methods. The DFT wavefunctions were used as trial wavefunctions for the single-determinant QMC calculations. Additionally, the one- and two-electron integrals generated during the DFT calculations were utilized as inputs for the selected Configuration Interaction (sCI) calculations, ensuring consistency across all three levels of theory.

\subsection{\textbf{Quantum Monte Carlo}}
QMC techniques are stochastic methods that solve the many-body Schrödinger equation with high accuracy using a limited but controlled number of approximations. By explicitly including many-body electronic interactions, these methods achieve chemical accuracy while maintaining favorable scaling compared to traditional high-level quantum chemistry approaches.
In this work, we leverage the extensive refactoring of QMCPACK that has been specifically optimized for exascale computing platforms over the past two decades. Building on previous optimizations that targeted material science applications, we extended and enhanced the Linear Combination of Atomic Orbitals (LCAO) branch of QMCPACK with new optimizations and data layouts specifically designed for GPU acceleration in molecular applications. These enhancements enable unprecedented scaling on Aurora's architecture for complex molecular systems.
Among the various QMC approaches, we employed Diffusion Monte Carlo (DMC) due to its ability to recover a significant portion of the correlation energy while maintaining favorable scaling with system size. Importantly, DMC operates directly in real space, meaning it does not rely on a finite basis set. As a result, DMC is free from basis set incompleteness error and can be viewed as operating in the infinite-basis limit by construction.  DMC solves the Schrödinger equation in imaginary time  $\tau= it$,  ensuring that any initial state $|\psi\rangle$  not orthogonal to the ground state $|\phi_0\rangle$ will converge to the ground state in the long-time limit:

\begin{equation}
\lim_{\tau \rightarrow \infty} \Psi(\textbf{R},\tau)=c_0 e^{-\epsilon_0\tau}\phi_0(\textbf{R})
\end{equation}

To address the fermion sign problem, we employed the fixed-node (FN) approximation\cite{Anderson1980}, which constrains the nodal surface of the many-body wavefunction to match that of a trial wavefunction. This introduces the only systematic error in DMC when the reference wavefunction is not exact. To minimize this error, we used a trial wavefunction of the Slater-Jastrow form:
\begin{equation}
\Psi_T(\vec{R}) = \exp\left[\sum_i J_i(\vec{R})\right]\sum_k^M C_kD_k^{\uparrow}(\phi)D_k^{\downarrow}(\phi)
\end{equation}
where $D_k^{\uparrow}(\phi)$ and $D_k^{\downarrow}(\phi)$
are Slater determinants for up and down spin electrons expressed in terms of single-particle orbitals $\phi_i = \sum_l^{N_b} C_l^i \Phi_l$ 
 The Jastrow factor $\exp[\sum_i J_i(\vec{R})]$
explicitly accounts for electron-electron correlations and significantly improves the trial wavefunction quality while maintaining the original nodal structure.

For our QMC calculations, we used the QMCPACK code\cite{QMCPACK1,QMCPACK2,QMCPACK3,QMCPACK4}, which has been specifically optimized for exascale computing platforms. The single-particle orbitals in our trial wavefunctions were generated using the $\omega$B97M-D3BJ functional as implemented in the PySCF package. This range-separated hybrid meta-GGA functional with D3BJ dispersion corrections was chosen for its excellent balance between accuracy and computational efficiency for organic and biological systems. All calculations employed correlation-consistent effective core potentials (ccECP) to reduce computational cost while maintaining high accuracy for valence electrons.

For each molecular system, we optimized up to 40 variational parameters, including one-body, two-body, and three-body Jastrow factors, using a variant of Umrigar's linear method. This optimization was performed using variational Monte Carlo (VMC) before the subsequent DMC calculations. The Jastrow parameters were optimized independently for each molecule to ensure optimal trial wavefunction quality. All DMC calculations used a time step of 0.01 a.u., which we verified through time-step extrapolation studies to be within error bars of the zero time-step limit. To ensure statistical independence and proper convergence, we used 4096 walkers per calculation, distributed across the exascale computing resources.\\
The stochastic nature of QMC offers exceptional parallel efficiency on modern supercomputers, with computational tasks being evaluated independently across thousands of processing units. This "embarrassingly parallel" characteristic positions QMC as an ideal candidate for exascale applications, allowing us to simulate systems containing thousands of electrons while maintaining quantum mechanical accuracy.

\subsubsection{\textbf{QMC Energy and Wavefunction Evaluation} \label{sec:qmc_energy}}
\begin{itemize}
    \item \textbf{Summary of Contributions}
\end{itemize}
We developed a highly optimized implementation of QMC calculations in QMCPACK that leverages the massive parallelism of exascale hardware. Building on previous optimizations for material science applications, we specifically enhanced the Linear Combination of Atomic Orbitals (LCAO) branch to accelerate molecular applications on GPU architectures. Our implementation enables unprecedented calculation throughput, processing over 40,000 molecular configurations with QMC (20,251 with forces) on the Aurora supercomputer, achieving 124 PFLOPS sustained performance on 2048 nodes, and while using over 4500 walkers per rank (see Fig\ref{fig:qmc_water_clusters}). 
\begin{itemize}
    \item \textbf{Algorithmic Innovations}
\end{itemize}

The core innovation in our QMC implementation is a multi-level batching approach that maximizes GPU utilization and minimizes kernel launch overhead:
\textbf{Walker-Level Batching} Multiple Monte Carlo walkers are evaluated simultaneously, transforming independent stochastic samples into batched operations that better utilize GPU parallelism.\\
\textbf{Multi-Center Batching} Rather than evaluating atomic orbitals one center at a time, we group centers by species and process them in batches. This approach:
\begin{itemize}
    \item Flattens the combined dimension of \{centers, electrons\} into a single array dimension
    \item Groups centers by species to minimize kernel launches Maintains consistent indexing schemes for reading and writing data
    \item Accumulates all partial orbital contributions in a minimal number of passes.
\end{itemize}

\textbf{Unified Data Layout}: We implemented consistent flattening schemes that eliminate index mismatches between writing and reading orbital data, preventing out-of-bounds memory accesses and simplifying debugging.\\
\textbf{Optimized Kernel Structure}: After computing radial (Rnl) and angular (Ylm) expansions in separate kernels, we combine partial results in a single kernel that maximizes GPU occupancy.

These implementations reduced kernel launch overhead by an order of magnitude and significantly improved memory throughput, which is particularly important for the LCAO approach used in molecular systems where the number of electrons is small. The GPUs' architecture-independent optimizations achieved 60\% of theoretical peak performance for typical molecular configurations in our dataset. 

\subsubsection{\textbf{Single Determinant QMC Forces }\label{sec:qmc_force}}
\begin{itemize}
    \item \textbf{Summary of Contributions}
\end{itemize}
For the QMC layer of our quantum chemistry database generation protocol, we build upon the existing force evaluation formalism in QMCPACK, developed according to Filippi et al.'s ~\cite{Filippi2016} approach. Rather than re-implementing the mathematical framework, we focused on optimizing the computational efficiency for exascale architectures by integrating the force evaluation with QMCPACK's batched structure. This optimization enabled us to compute accurate interatomic forces for over 20,000 molecular configurations, achieving 62 PFLOPS sustained performance on Aurora with scaling efficiency of 30\% up to 2048 nodes.

By optimizing the force evaluation code to utilize the same batching structures we developed for energy calculations, we achieved an acceleration of 12$\times$ compared to the original implementation while maintaining the same statistical accuracy. This performance gain removes a critical bottleneck in using QMC for large-scale force calculations, enabling the largest database of QMC-Forces for machine learned potentials. Additional components of the force evaluation remain on the CPU and lead to significant data transfers hindering performance. Future developments will focus particularly on porting them to GPU.
\begin{itemize}
    \item \textbf{Algorithmic Innovations}
\end{itemize}
Our primary innovation was adapting the existing force evaluation implementation to utilize the batched structure we developed for wavefunction evaluation:\\
\textbf{Efficient Matrix Formalism}: We implemented the complete mathematical framework from Filippi et al., which expresses derivatives and one-body operators through a unified approach. The logarithmic derivative of a Slater determinant with matrix $A_{ij} = \phi_j(r_i)$, the logarithmic derivative with respect to any parameter $\lambda$ is:

\begin{equation}
\frac{d \ln D}{d\lambda} = \text{tr}(A^{-1}B) \quad \text{where} \quad B = \frac{dA}{d\lambda}
\end{equation}
This formulation allows us to compute forces as: 
\begin{equation}
\frac{\partial E_L}{\partial R_a} = \text{tr}\left(A^{-1}\frac{\partial B}{\partial R_a} - X\frac{\partial A}{\partial R_a}\right)
\end{equation}

where $X = A^{-1}BA^{-1}$. The derivatives of the matrices $A$ and $B$ with respect to nuclear coordinates are fully implemented, including the complex expressions for the non-local pseudopotential contribution. This implementation verified that the computational cost scales as $O(N^3*3*N_{atom})$ with minimal overhead compared to energy-only calculations.

\textbf{Integration with Multi-Walker Batching}: We restructured the force evaluation code to process multiple walkers simultaneously. Rather than independently computing forces for each walker, we leveraged the batched walker approach to transform the calculations into batched matrix operations. This significantly improved GPU utilization and reduced kernel launch overhead.\\
\textbf{Extending Multi-Center Batching to Forces}: We extended our multi-center batching approach from energy calculations to force evaluations. This required careful adaptation of the force calculation kernels to process multiple atomic centers in batches, maintaining the same flattened indexing scheme used in the wavefunction evaluation. Each atomic derivative contribution is now computed in batches grouped by atomic species, dramatically reducing the number of kernel launches.\\
\textbf{Unified Memory Management}: We implemented an optimized memory management scheme that reuses the data structures from the energy evaluation for force calculations. This approach minimizes memory transfers between CPU and GPU and ensures consistent data access patterns across different parts of the calculation.\\
\textbf{Optimized Space-Warp Implementation}: We adapted the space-warp coordinate transformation to our batched structure. The transformation, which reduces statistical variance in force estimators, was implemented with vectorized operations that process multiple walkers and centers simultaneously, improving computational efficiency.

By integrating force calculations with the batched structure of QMCPACK, we eliminated redundant computations and maximized hardware utilization, enabling force calculations at scales previously impossible with QMC methods.
The performance improvements from our batched implementation make QMC force calculations practical for large-scale applications, opening new possibilities for highly accurate MD simulations and foundation model training based on QMC-level accuracy.

\subsubsection{\textbf{Multideterminant QMC Forces}}\label{sec:qmc-msd-forces}
\begin{itemize}
    \item \textbf{Summary of Contributions}
\end{itemize}
An important advancement in our work is the full implementation and integration of multideterminant QMC forces in QMCPACK. While some elements of the mathematical framework existed in previous versions, these implementations were incomplete and non-functional for practical calculations. We significantly extended, completed, and optimized these components, then fully integrated them with our batched computation structure to enable exascale performance. With this implementation, we successfully computed forces for 2,000 molecular configurations using multideterminant QMC with selected Configuration Interaction (sCI) wavefunctions, achieving 52 PFLOPS on Aurora, using 200 molecules per run and 20 nodes per molecule, with 1.4M determinants per molecule.
This represents a meaningful methodological advancement, as multideterminant approaches provide a systematically improvable path toward more accurate solutions of the many-body Schrödinger equation by reducing the fixed-node error that limits single-determinant QMC. Prior to our work, such calculations remained largely theoretical or limited to small proof-of-concept systems.

\begin{itemize}
    \item \textbf{Algorithmic Innovations}
\end{itemize}
Our implementation required completing and integrating several key components of the multideterminant QMC force framework:\\
\textbf{Completion of Mathematical Framework}: Building on Filippi et al.'s theoretical approach, we completed the implementation of the excitation-based formalism for evaluating multideterminant forces. For a multideterminant wavefunction we finalized all necessary components to evaluate derivatives of ratios of sums of determinants, debugging and extending the existing code to handle excited determinants properly.\\
\textbf{Integration with Batched Architecture}:We restructured the multideterminant force evaluation to leverage our batched computation framework. This integration ensures optimal performance on exascale hardware by:
\begin{itemize}
    \item Processing multiple walkers simultaneously
    \item Grouping determinants by excitation level for batch processing
    \item Utilizing the same flattened center indexing scheme used elsewhere in the code.
\end{itemize}

\textbf{Optimization of Table Method Implementation}: We refined and optimized the table method algorithm for evaluating excited determinant quantities within our batched framework. This approach now efficiently handles the complex data structures required for multideterminant calculations. Our implementation of multideterminant force calculation re-uses many of the quantities that are already required for the single-determinant force calculation and adds a cost that scales as $O(N_{det}*3*N_{atom})$.\\
\textbf{Improved Memory Management}: We developed enhanced memory management schemes that:
\begin{itemize}
    \item  Minimize data transfers between CPU and GPU;
    \item Ensure optimal data locality for GPU processing;
    \item Reuse memory allocations across different parts of the calculation.
\end{itemize}

\textbf{Integration with sCI Workflow}: We established a computational pipeline that connects our selected Configuration Interaction (sCI) implementation with the multideterminant QMC force evaluation. This integration enables the automatic generation, optimization, and force evaluation of multideterminant wavefunctions in a cohesive workflow.

This implementation makes multideterminant QMC forces practically usable at scale for the first time. By completing, optimizing, and integrating this capability with our batched computation structure, we have created a pathway for generating higher-accuracy reference data for ML potentials, particularly for systems where electronic correlation effects are significant.
\subsubsection{\textbf{Selected Configuration Interaction}}
\begin{itemize}
    \item \textbf{Summary of Contributions}
\end{itemize}
\
The third layer of our quantum chemistry database generation protocol employs selected Configuration Interaction (sCI) methods, which provide a systematically improvable path toward the exact solution of the many-body Schrödinger equation. Our sCI implementation is based on the CIPSI (Configuration Interaction using a Perturbative Selection done Iteratively) algorithm, which we implemented in C++ and ported to GPUs to address Exascale computing.

Our sCI code is only several months old and is still undergoing rapid development. There are many more algorithmic improvements underway and planned for the near future, but it is already capable of leveraging GPUs, and for this work it was able to generate 1.4M-determinant wavefunctions for 2000 molecules using approximately 1k node-hours, which represents less than 1\% of the total computational cost of the full  {\{DFT\textrightarrow sCI\textrightarrow VMC/DMC\}} workflow per molecule. So even at a very early stage of development, with many planned improvements that will increase the capability of the code, we are still able to greatly improve our nodal surface quality (by adding multideterminant expansions) without adding any significant cost in performing the sCI calculations. Crucially, because we ultimately project these wavefunctions with DMC, we eliminate the limitations of the finite basis used in sCI. The final energies and forces are obtained in real space and are thus free from basis set incompleteness error, making the approach both efficient and systematically improvable.

\begin{itemize}
    \item \textbf{Algorithmic Innovations}
\end{itemize}

The core of our implementation follows an iterative procedure in which we begin from an eigenvector of the Hamiltonian in a given set of determinants, and we repeat the following steps until some convergence criterion is met:
\begin{enumerate}
\item Generate all single and double excitations from the current space
\item Compute second-order perturbative energy contributions
\item Select determinants with the largest contributions
\item Diagonalize the Hamiltonian in the new variational space
\end{enumerate}

The C++ implementation has been specifically optimized for exascale computing with a focus on efficient memory management and parallel execution. Key optimizations include:
\begin{itemize}
\item Bit-encoded determinant representation using native fixed-width integer types for compact storage and fast comparison operations
\item Task-based parallelism for generation and evaluation of candidate determinants
\item GPU-accelerated Hamiltonian diagonalization (Davidson-Liu)
\end{itemize}

For force evaluation with sCI wavefunctions, we leverage our QMC implementation rather than developing a separate analytical gradient procedure. This approach offers several advantages: it avoids the complication of computing analytical derivatives of the sCI wavefunctions (which would require more input data than the standard FCIDUMP interface provides), it naturally incorporates the Jastrow factor which significantly improves the wavefunction quality, and it provides a unified computational framework for energy and force evaluation. The multideterminant wavefunctions generated by sCI are directly imported into QMCPACK, where they serve as a starting point for further optimization before force evaluation.

This integration of sCI and QMC enables an efficient workflow where high-quality nodal surfaces are generated through the sCI procedure, then refined and used for force evaluation in QMC, circumventing many of the challenges associated with analytical gradients for large multideterminant expansions while maintaining high accuracy.

\subsubsection{\textbf{QMC Submission Engine on Aurora}}

QMCPACK calculations on Aurora were facilitated by a custom two-level job submission engine built on top of the Nexus\cite{krogel2016} workflow system. To enable to maximize computational efficiency, molecular systems were batched together by electron count similarity as this supports greater execution uniformity among batched molecules.  This top level parallelism, across batches, was complemented by further separation of molecules into sub-ensembles, each of which ran under a single top-level QMCPACK MPI process to maximize robustness of the simulations, providing a simple form of fault tolerance.  The balance between these levels, as well as assignment of individual molecules to selected node resources, was optimized to simultaneously maximize job submission size, fault tolerance, and computational efficiency.
\section{Performance Measurement}
To demonstrate the computational efficiency and accuracy of our new QMC implementation, we performed extensive performance analyses on several representative molecular systems. We primarily focused on water molecules (monomer, dimer, and trimer) due to their biological relevance and the well-known computational challenges they pose for accurate quantum chemical treatment. Additionally, we benchmarked a representative SPICE molecule with heavier atoms containing 16 electrons for force evaluation, as well as a larger, computationally demanding molecule with 176 electrons to examine scalability.\\

\begin{figure}
    \centering
     \includegraphics[width=0.45\textwidth]{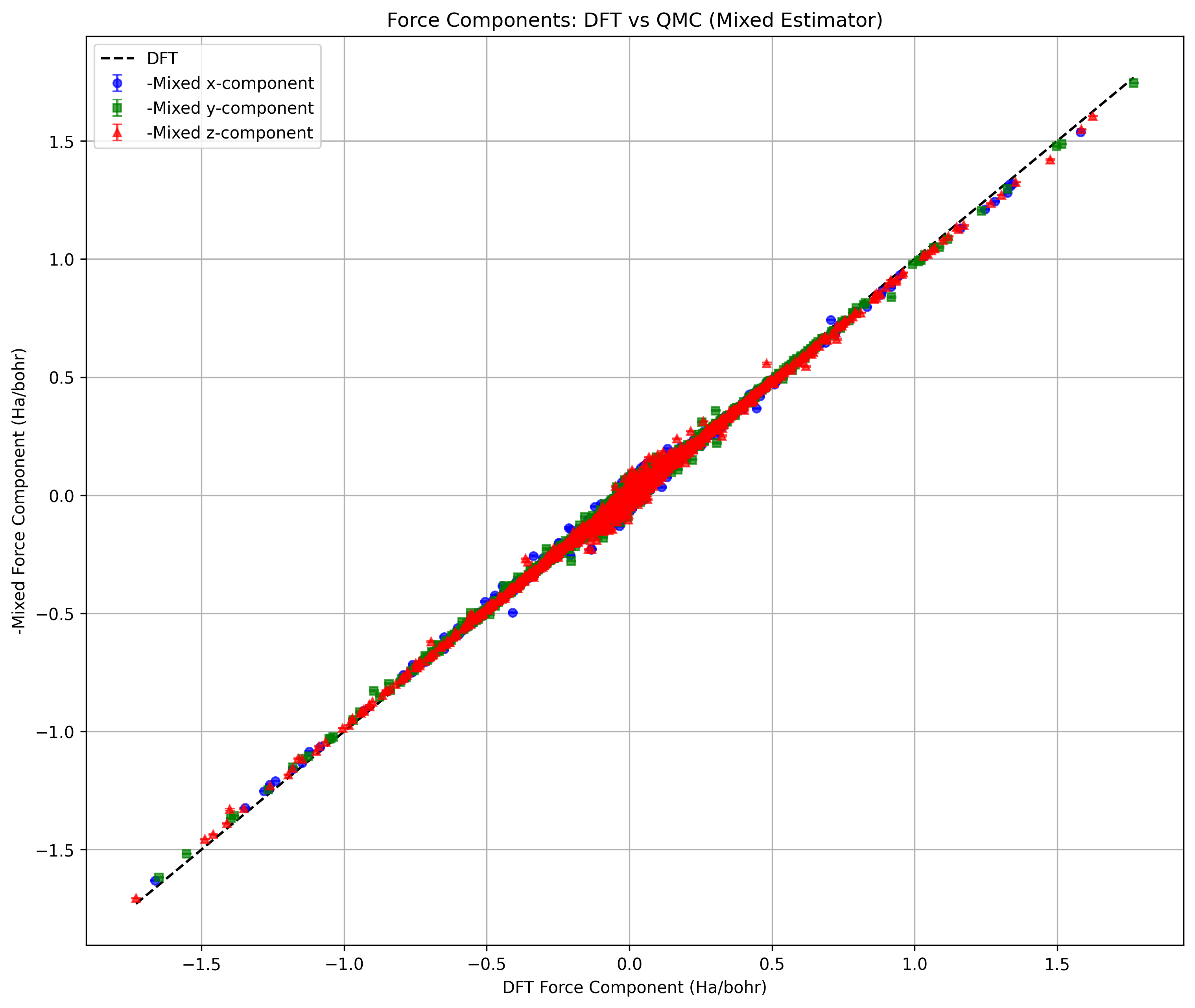}
        \caption{Comparison of force components between DFT and QMC mixed estimator. The three different marker shapes and colors represent the Cartesian components of the forces. The forces correlate with the DFT very well while the DFT shows significant discrepancies due to the lack of accuracy of the method.}
        \label{fig:atomization}
   \label{fig:force_energy_comparison}
\end{figure}

Figure \ref{fig:force_energy_comparison} presents a comprehensive comparison between the DFT forces and the QMC mixed estimator forces for a representative dataset. In these plots, the x-axis represents the DFT force components (in Ha/bohr), while the y-axis corresponds to the mixed estimator forces, also in Ha/bohr which is computed via the mixed estimator defined as $F{mixed}=2F{DMC}-F{VMC}$. Across the three Cartesian components (x, y, z), a strong linear correlation is observed between the DFT forces and the QMC mixed estimator $F_{mixed}$ as demonstrated by the clustering of points along the ideal diagonal line (shown as a dashed line in each subplot). The slight systematic differences are present between the DFT values and the Mixed estimator forces show the subtle differences between the methods and reflecting the increase of accuracy.

\section{Performance Results}
\subsubsection{\textbf{Single-Determinant QMC Performance}}

\begin{figure}[htbp] \centering \includegraphics[width=0.45\textwidth]{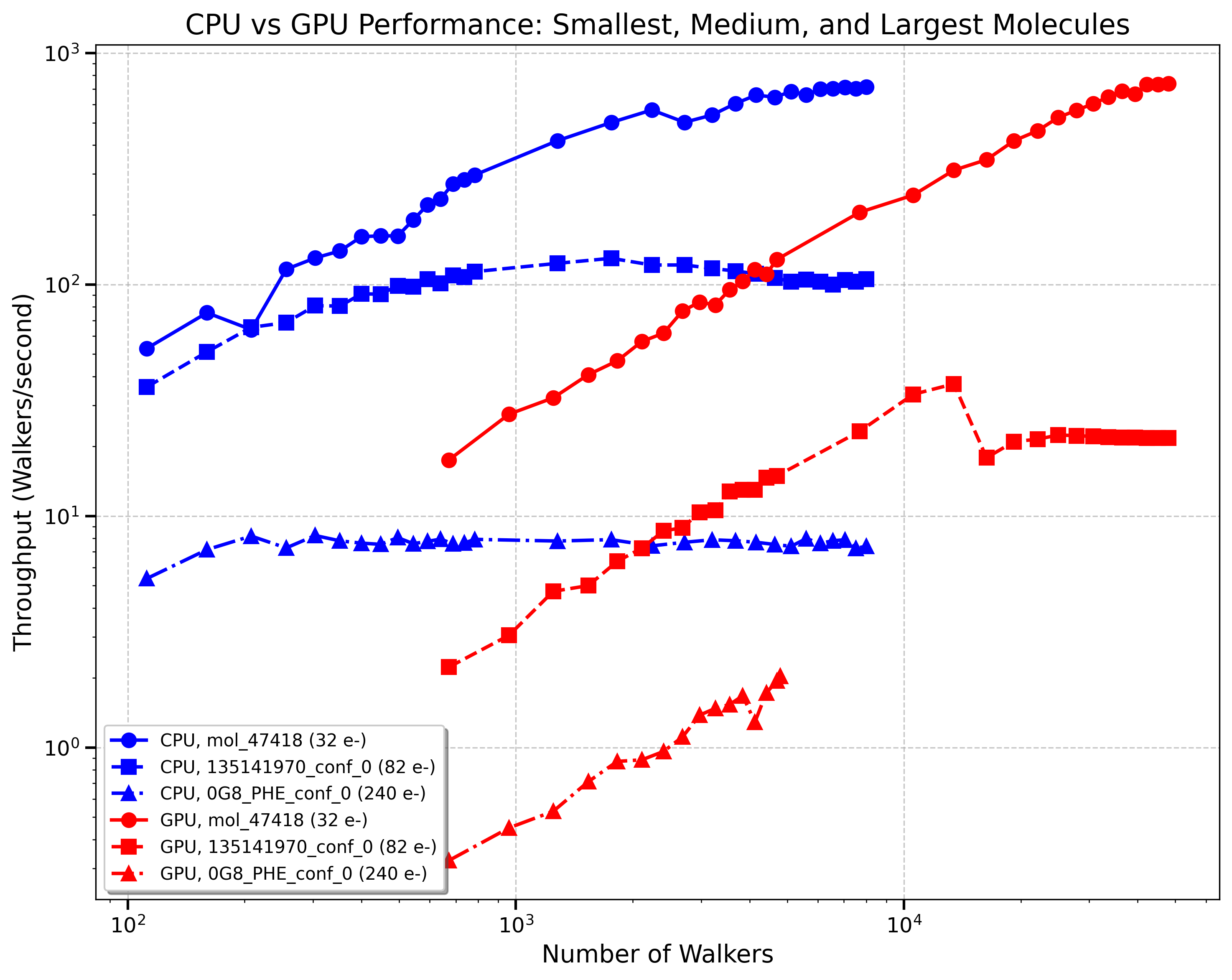} \caption{Performance comparison of the initial QMC algorithm on CPU versus GPU for three representative molecular systems with different electron counts (32, 82, and 240 electrons). The CPU implementation consistently outperformed the GPU version, highlighting limitations in data layout, memory overhead, and kernel-launch inefficiencies in the original GPU algorithm.} \label{fig:qmc_initial_performance} \end{figure}

Our initial performance analysis revealed a counterintuitive finding: the original GPU implementation of the QMC algorithm significantly underperformed compared to the CPU-only implementation, particularly for molecular systems with electron counts common in biological applications (Figure \ref{fig:qmc_initial_performance}). As shown, the CPU version consistently delivered higher throughput for molecular systems containing 32, 82, and 240 electrons. The GPU implementation only outperformed the CPU in extremely high walker-count scenarios, a regime practically limited by memory saturation and therefore infeasible for routine calculations.

This stark performance gap highlighted the need for fundamental algorithm redesign, specifically addressing GPU kernel overhead, parallel efficiency, and memory optimization as discussed in Sections \ref{sec:qmc_energy} and \ref{sec:qmc_force}.

\begin{figure}[htbp] \centering \includegraphics[width=0.45\textwidth]{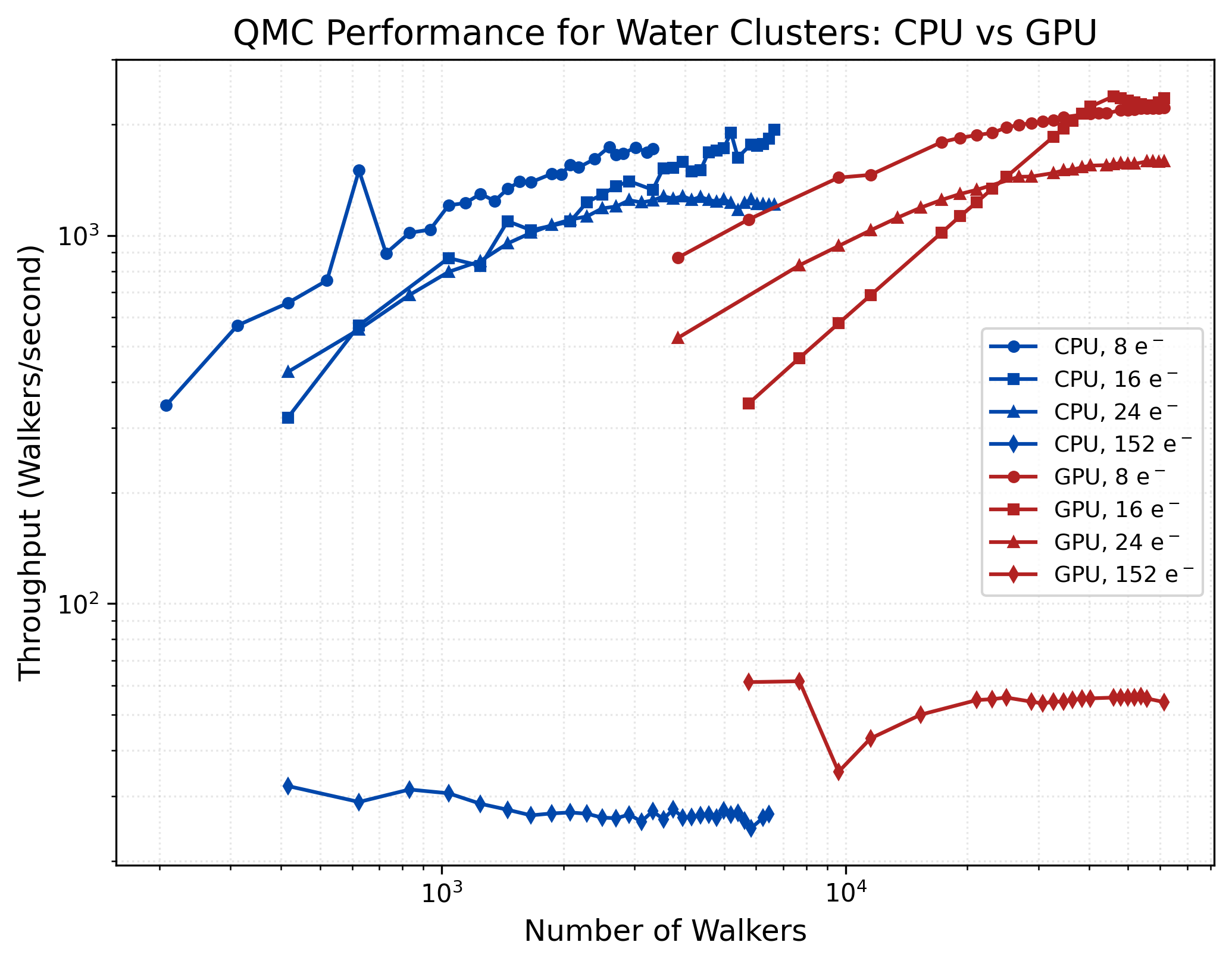} \caption{Performance comparison of CPU versus optimized GPU implementations for QMC calculations on water clusters with varying electron counts (8, 16, 24, and 152 electrons). Throughput is shown as walkers per second versus total walker count, demonstrating robust GPU acceleration and scalability.} \label{fig:qmc_water_clusters} \end{figure}

Figure \ref{fig:qmc_water_clusters} illustrates the substantial performance gains achieved with our new algorithm. We observe clear improvements across water molecules and clusters with electron counts ranging from 8 to 152 electrons. Quantitative speedups measured against the CPU baseline are summarized in Table \ref{tab:qmc_gpu_speedup}, highlighting consistent and significant GPU acceleration.

\begin{table}[htbp]
\centering
\caption{Measured speedups for QMC energy evaluation on GPUs compared to CPUs, utilizing the optimized algorithm.}
\label{tab:qmc_gpu_speedup}
\begin{tabular}{cc}

\toprule
\textbf{Electron Count} & \textbf{Speedup (GPU vs. CPU)} \\
\midrule
8 & 1.29 \\
16 & 1.22 \\
24 & 1.31 \\
152 & 2.01 \\
\bottomrule
\end{tabular}
\end{table}

The significant GPU acceleration is due to our multi-level batching approach, optimized data layouts, and careful tuning of the number of walkers per GPU rank, maximizing occupancy and minimizing kernel overhead. Because QMC methods are embarrassingly parallel, increasing the number of walkers per rank enabled higher throughput with fewer overall simulation steps, thus effectively utilizing GPU resources.

\subsubsection{\textbf{Resource Allocation Strategy}}

Through detailed performance analyses, we established an optimal resource allocation strategy that balances computational efficiency with statistical accuracy requirements. As QMC is an embarrassingly parallel method, we calibrated our runs to achieve a fixed statistical error target of approximately 0.0005 mHa (below the chemical accuracy threshold of 1 kcal/mol). Following standard Monte Carlo statistics, the error bars scale as $1/\sqrt{N_{\text{samples}}}$, requiring us to carefully balance the number of walkers per rank with the total number of sampling steps.

To optimize throughput while maintaining accuracy, we determined that the ideal configuration was 4320 walkers per GPU rank. We then scaled the number of nodes based on system size to achieve the desired statistical convergence within reasonable wall-clock times compatible with supercomputer queue policies. This translated to allocating two Aurora nodes per small molecule (up to 16 electrons), six nodes for medium-sized systems like the water trimer (24 electrons), and correspondingly more for larger systems.

For force calculations, which are approximately 100$\times$ more computationally intensive than energy-only evaluations due to known bottlenecks in the current implementation (e.g. finite differences for space warp transformation), this scaling approach was particularly critical. Previous QMC implementations were simply incapable of executing large-scale force evaluations, making direct comparative benchmarks unavailable. Our batched implementation made these calculations feasible within our computational budget, enabling us to process approximately 20,000 molecular configurations with full force evaluation. Future optimization efforts will focus on further reducing GPU-to-host data transfers and increasing GPU kernel efficiency to improve both throughput and statistical precision.
\subsubsection{\textbf{Multideterminant QMC Performance}}

Multideterminant calculations drastically increase computational demands compared to single-determinant approaches. In principle, multideterminant wavefunction evaluations scale as $\sqrt{N_{det}}$, where $N_{det}$ represents the number of determinants included and in this context, limited to 1.4M determinants leading to a PT2 of {\raise.17ex\hbox{$\scriptstyle\sim$}}2 mHa. This characteristic scaling implies that achieving the highest accuracy in multideterminant calculations rapidly becomes computationally prohibitive as the number of determinants increases.

We successfully executed multideterminant QMC calculations for the water monomer; however, performing analogous calculations on the water dimer exceeded our computational allocation limits, underscoring the extreme intensity of such computations. To achieve optimal performance with multideterminant wavefunctions, we allocated approximately 20 Aurora nodes per molecule and processed batches of 100 molecules per job (utilizing 2000 nodes per job submission). Our implementation reached a sustained computational throughput of 24 TFLOPS per node configuration on Aurora, corresponding to 25\% of the peak theoretical performance.
These substantial accuracy gains justify the computational costs incurred, demonstrating the critical importance of multideterminant QMC methods in achieving chemically accurate reference datasets for training next-generation ML models.

\section{Implications}
We presented an end-to-end integrated computational strategy for generating synthetic quantum chemistry datasets that significantly enhance the accuracy and predictive power of atomistic Foundation ML models. Our method pushes the boundaries of quantum chemical calculations, especially Diffusion QMC and selected Configuration Interaction, to unprecedented scales. Leveraging the power of exascale computing resources, we computed energies and forces with near-exact quantum accuracy at the CBS limit for tens of thousands of molecular configurations, establishing a new state-of-the-art in dataset generation. 

Key algorithmic and computational innovations include a highly optimized multi-level batching scheme, a novel unified memory management strategy, and the optimization for GPUs of zero-variance force estimators within the QMC method. Together, these advancements enabled efficient and scalable calculations on leadership-class computing facilities such as Aurora. Specifically, our optimized GPU implementations achieved unparalleled performance, reaching sustained multi-petaflops computational throughput and overcoming significant bottlenecks traditionally associated with QMC force computations. Improving further the accuracy beyond the present DMC and the combination of multi-determinant QMC energies and forces with selected-CI wavefunctions could be difficult with classical computers due to the previously discussed Exponential Wall problem. However, once logical qubits become available\cite{zhang2025fault}, Quantum Computing could be a solution to overcome this barrier and bring, at least, useful polynomial speedups\cite{lee2023evaluating}. As this would allow access to exact Schrödinger-like Full-CI/CBS computations (Figure 1), we already considered these new class of quantum algorithm for chemistry in the construction of our present dataset. As an Easter egg, we included a dozen datapoints, converged at the CBS chemical accuracy that were obtained with our GPU-accelerated quantum emulator\cite{traore2024shortcut}. This should generalize in the future as the algorithms, quantum emulators and quantum hardware mature.
Overall, we demonstrated the practical applicability of our computational pipeline using the FeNNix-Bio1 foundation NNP model trained with our high-fidelity datasets. Therefore large scale applications became possible.
\subsection{\textbf{Large Scale Quantum Dynamics Simulations with a "beyond DFT" Foundation Neural Network Model}}
\begin{itemize}
    \item \textbf{Summary of Contributions:} 
    \end{itemize}
Using the CBS QMC and sCI data produced in the project, we improved an early version of the FeNNix-Bio1 model \cite{FennixBio} using transfer ML via the FeNNol library\cite{ple2024fennol} to obtain a "beyond DFT" improved approach. We then coupled the improved model to the GPU-accelerated massively parallel Tinker-HP MD package\cite{lagardere2018tinker,adjoua2021tinker} and leverage simultaneously its Deep-HP (neural networks), Quantum-HP (quantum nuclear effects) modules while using enhanced sampling via non-supervised adaptive sampling MD.
We tested the new approach on both Jean-Zay (H100) and Aurora(Xe) machines on the full structure of the Satellite Tobacco Virus, a 1-million atom virus. The simulation encompassed 32-million particles (1-million atom X 32 beads) since it leverages Ring
Polymer quantum MD simulation. We extended the conformational space sampling with  adaptive sampling replica to decompose the global exploration problem into a set of separate MD trajectories. These latter can be restarted thanks to  a selective process in order to achieve sufficient phase-space sampling. \\

\begin{figure}[htbp] \centering \includegraphics[width=0.45\textwidth]{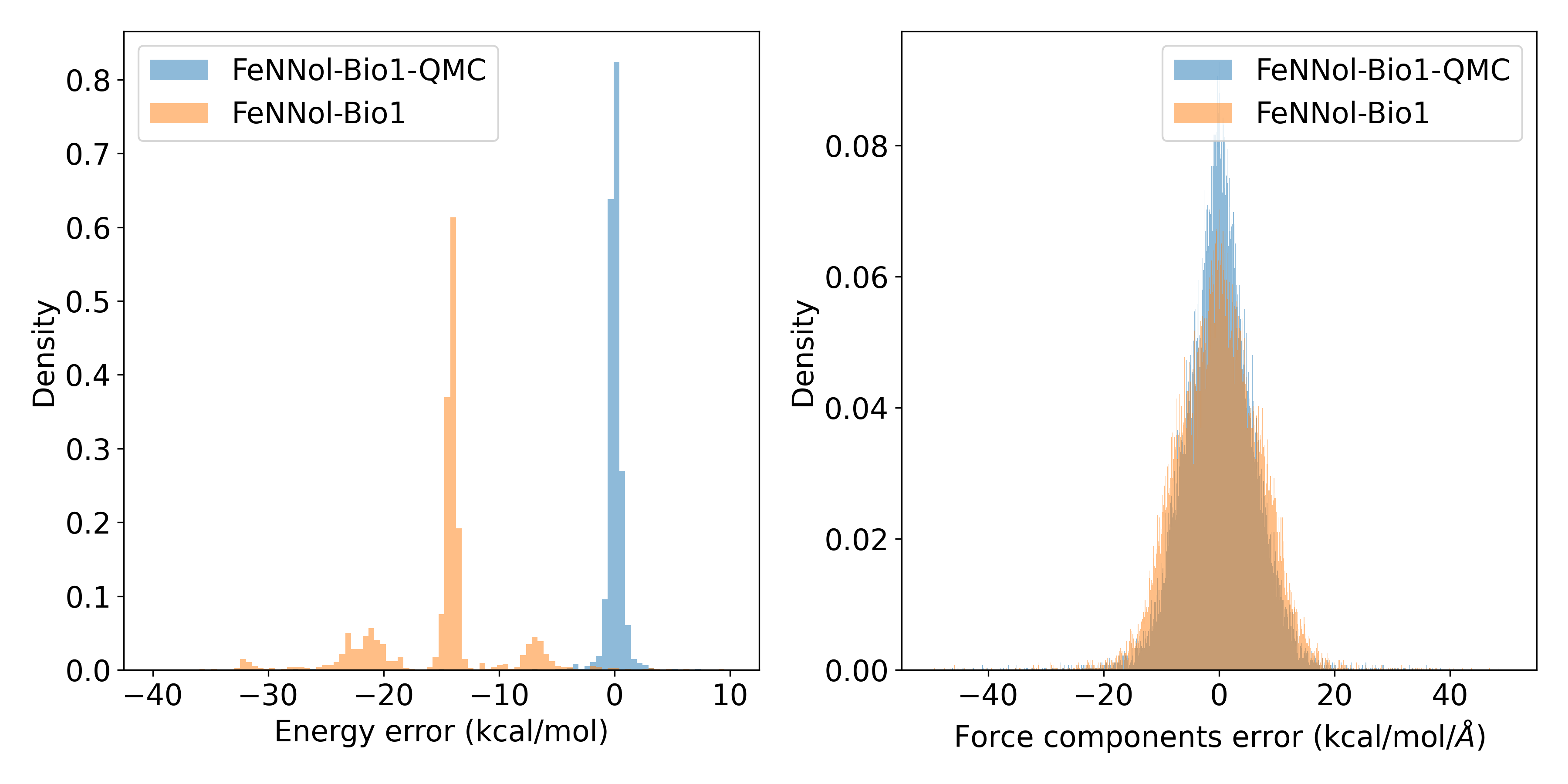} \caption{Distribution of the energy and force components errors on the QMC validation set, before and after transfer learning} \label{fig:fennol_bio1_QMC_histograms} \end{figure}

\textbf{Algorithmic Innovations}
\begin{itemize}

\item{\textbf{Leveraging Transfer Learning to Incorporate High Accuracy Quantum Chemistry Data into FeNNix-Bio1}}
\end{itemize}
As a proof of concept, in order to improve further the quality of the DFT-based FeNNix-Bio1 model towards a "beyond DFT" model, we leveraged transfer learning,\cite{smith2019approaching} to include the high-quality, CBS, DMC energies and forces present in the SPICE2(+)-ccECP dataset. In practice, starting from the model trained on the DFT part of the dataset, we reoptimize the energy heads (freezing the parameters of the embedding), on the QMC energies and forces with the same loss function as in the original training. As is common practice, we started with a smaller learning rate of $10^{-5}$ using the adabelief optimizer\cite{zhuang2020adabelief}. 
Note that the QMC energies and forces inherently contain statistical noises, with a significantly larger variance for the forces. We take this into account in the loss function that we optimize for transfer learning by putting 100 times more weight on the energies than on the forces. 
Starting from a MAE of 15.3 kcal/mol (RMSE of 16.3 kcal/mol) on the energies and MAE of 5.7 kcal/mol/\AA\ (RMSE of 8.7 kcal/mol/\AA) on the forces, we end up, on the validation set, after 1000 epochs of training, with a MAE of 0.54 kcal/mol (RMSE of 2.4 kcal/mol) on the energies and MAE of 4.6 kcal/mol/\AA\ (RMSE of 6.3 kcal/mol/\AA) on the forces.
These results are shown in Figure \ref{fig:fennol_bio1_QMC_histograms}.

\begin{figure}[htbp] \centering \includegraphics[width=0.45\textwidth]{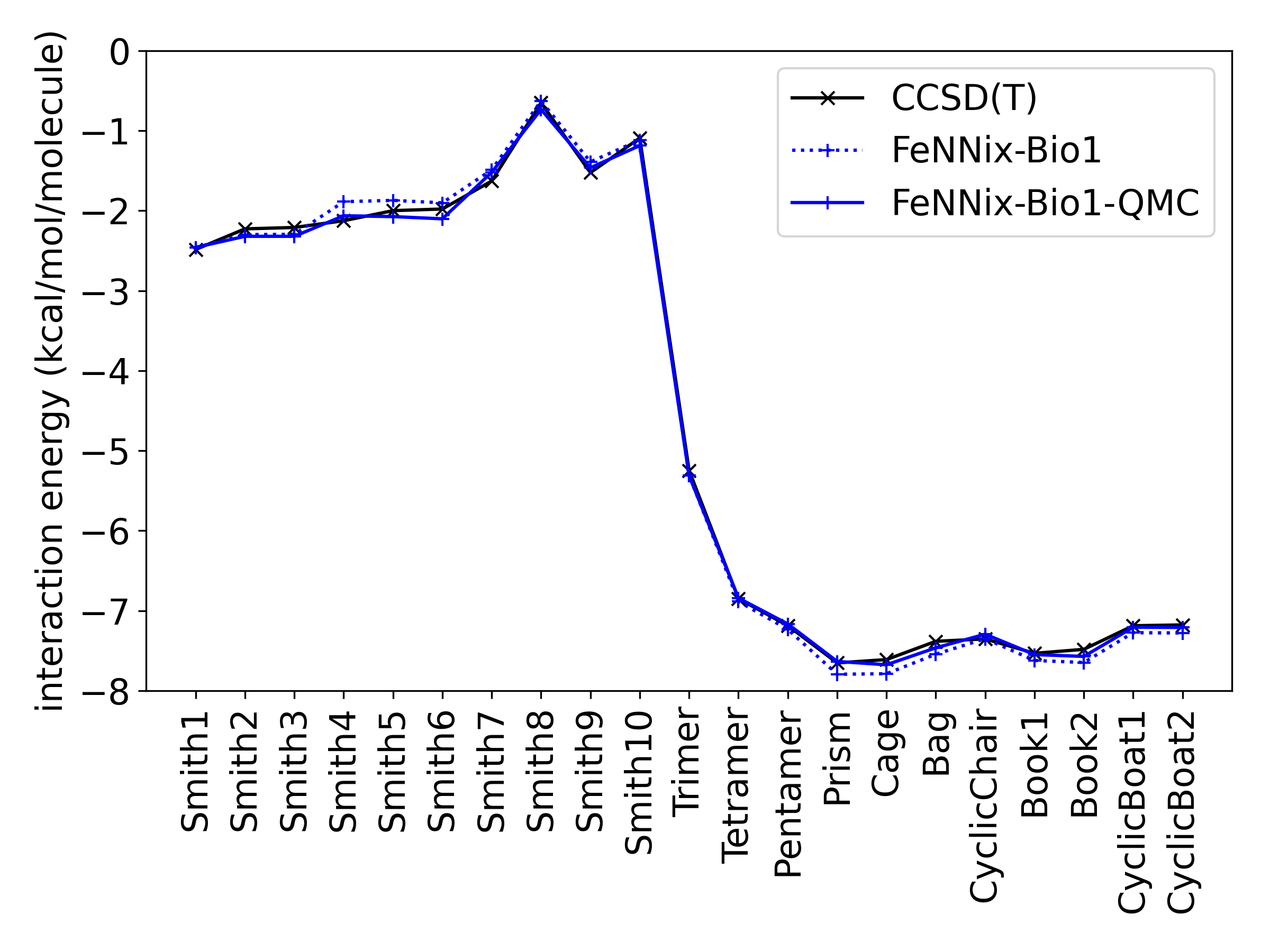} \caption{Per molecule interaction energy on the low-lying water clusters, before and after transfer learning, compared to CCSD(T)/CBS\cite{tschumper2002anchoring,bates2009ccsd}} \label{fig:water_clusters} \end{figure}

We assess the changes in the model due to the transfer learning by computing the interaction energies of low-lying water clusters up to hexamers\cite{tschumper2002anchoring,bates2009ccsd}. Compared to CCSD(T)/CBS energies, we observe a systematic improvement of the machine learning model as shown in Figure \ref{fig:water_clusters}. Starting from a MAE of 0.38 kcal/mol (RMSE of 0.5 kcal/mol), we end up with a MAE of 0.20 kcal/mol (RMSE of 0.24 kcal/mol). 

Beside accumulating more QMC data to better cover chemical space, future work will explore how to exploit the variance estimation provided by QMC in the transfer learning loss function.

As a note, it is important to quote that thanks to the generated dataset, the foundation model is not the only product that can be directly generated. Indeed, thanks to directed graph attention networks\cite{ChenGaff}, the discussed FeNNol library can generate, within hours, new version of classical force fields such as GAFF (Generalized Amber Force Field)\cite{wang2004development} that can be used to prepare initial simulation setups at very limited computational cost.

\begin{itemize}

\item \textbf{Bridging a Foundation Model to Ring Polymer Quantum Molecular Dynamics and to Adaptive Sampling}
\end{itemize}
We coupled our newly developed JAX-based \cite{jax2018github} FeNNoL library \cite{ple2024fennol} that allows for the construction and inference of atomistic ML potentials such as FeNNix-Bio1\cite{FennixBio} to the Tinker-HP MD package\cite{lagardere2018tinker,adjoua2021tinker} via the Deep-HP interface\cite{inizan2022scalable}. By doing so, we leverage the Ring Polymer MD (RPMD)\cite{habershon2013ring} capabilities of Tinker-HP (Quantum-HP module \cite{ple2022routine}) in combination with FeNNix-Bio1 in order to recover quantum dynamics and statistics of the Satellite Tobacco Virus (STMV)\cite{blazhynska2025satellite}. STMV is a good model system for studying viral assembly and structural dynamics as it is includes a single-stranded RNA genome located within an icosahedral capsid. RPMD can be used efficiently for determining accurate properties of condensed-phase molecular systems\cite{markland2018nuclear,mauger2021nuclear}. It uses classical statistical physics to mimic quantum dynamics via fictitious ring polymer which consist in n copies (or beads) of the system particles connected by harmonic springs\cite{ple2022routine}. Obtaining large scale reference simulation data using RPMD on meaningful biological systems will help to improve the present FeNNix-Bio1 near-classical computational cost quantum thermostats such as the Adaptative Quantum Thermal Bath\cite{mauger2021nuclear} that can already generate orders of magnitude longer simulations but that are more approximate. To ensure full accuracy at room temperature, we used 32 beads per atom in RPMD simulations, a number where simulations are essentially converged\cite{markland2018nuclear,mauger2021nuclear}. These latter leverage two levels of parallelism: one at the beads level and a second one within each bead. Additionally, to push further the sampling of the conformational space, we added on top of the RPMD framework an additional layer of parallelism using unsupervised adaptive sampling\cite{D1SC00145K}. Here multiple iterations of independent MD simulations replica are launched to "mine" the conformational space. The selection of the initial structures at each iteration follows an adaptive, fully automated, procedure enhancing the exploration of a low-dimensional space of slow variables and maximizing the conformational space exploration by penalizing areas that have already been extensively visited. This unsupervised selection step has the advantage of suppressing the problem of the choice of the initial collective variable at the beginning of the simulation. In practice the number of replica is only limited by the number of available GPUs, making adaptive sampling particularly suited for exascale computing\cite{D1SC00145K}. On Jean-Zay, each of the bead was parallelized on 8 H100 GPUs for a total of 8x32=256 H100s. Exploiting parallelism within each bead is mandatory to fit in the memory of each GPU thanks to memory distribution. All in all, we obtain 0.11 ns/day (mixed precision) per single adaptive sampling replica, making production of tens of nanoseconds per day of such quantum dynamics simulation accessible through adaptive sampling given enough resources. To our knowledge, achieving such level of quantum dynamics production with a foundation model grounded on extremely accurate quantum chemistry dataset on such a large biosystem is unprecedented.
The Jean Zay machine having 1280 GPUs, we were able to launch 4 simultaneous replica leveraging 1024 GPUs, producing therefore 0.44 ns/day of adaptive quantum dynamics. Note that such adaptive setup can also be used sequentially to extend the number of replica that can be launched one after another enabling to accumulate simulation time. Finally, further H100 optimization of the combined computing layers should possibly enable us to achieve the 1 ns/day threshold with the same number of replica whereas adding more replica is also a solution. This motivated our portage on the Exascale-class Aurora machine.

\begin{figure}[htbp] \centering \includegraphics[width=0.45\textwidth]{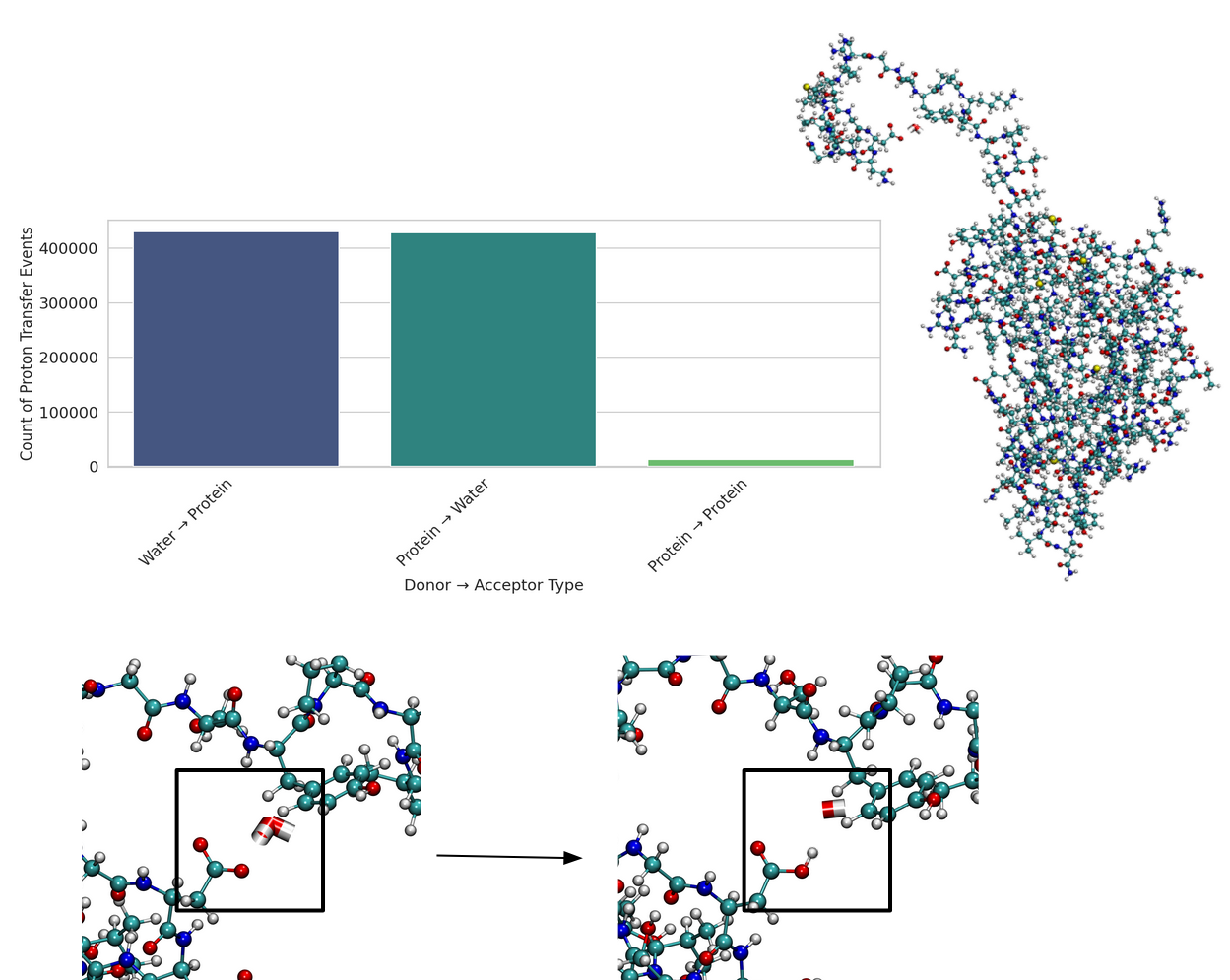} \caption{Number of proton transfer events among key donor–acceptor pairs: water–protein, protein–water, and protein–protein. The protein component corresponds to the STMV capsomer, illustrated to the right of the graph. The lower panel visualizations highlight a representative proton transfer event from a water molecule to a carboxyl group of a capsomer residue.} \label{fig:STMV_capsomer} \end{figure}

\begin{itemize}
\item \textbf{Aurora Portage of Tinker-HP/Deep-HP}
\end{itemize}
In the perspective to run Tinker-HP/Deep-HP (mainly written in Fortran 90) on Aurora, we have updated the NVIDIA-based GPU code from OpenACC/CUDA to OpenMP/SYCL. This direct strategy preserves the initial code structure, paths the way to portability across multi-platform, minimizes re-implementation effort, and technically keeps intact the global optimization. On one hand, the translation from OpenACC/CUDA to OpenMP/SYCL implementation preserves control over data transfers between CPUs and GPUs with minor exceptions, as well as parallelization of the computationally intensive kernels ; while on the other, distributed parallelism through the native 3D domain decomposition of Tinker-HP\cite{lagardere2018tinker,adjoua2021tinker} can be directly exploited through MPI calls between Intel Xe GPUs. Contrary to our simulations on Nvidia GPUs, we resorted to an Aurora-compatible version of Pytorch\cite{pytorchpaper} and not JAX for the computation of the NNP energy and forces. Following the same protocol described on Jean Zay (see previous section), we obtain 0.03 ns/day (mixed precision) per single adaptive sampling replica (32 nodes, 384 GPUs).As expected since this is a new portage of the code that will benefit of further optimization over the next months, the present production per replica on Aurora is lower than the numbers per replica obtained on Jean Zay which used the reference Tinker-HP implementation. One constraint is related to memory: Jean Zay's H100 propose 80Gb/GPU while Aurora is adding 64Gb per Xe GPU. However, Aurora benefits from its Exascale capabilities and from its associated extremely large availability of computing nodes. Aurora can achieve a similar performance as Jean Zay with 15 replica (5760 GPUs, 480 nodes).
\begin{itemize}
 \item \textbf{Large scale simulations on STMV: analysis of results}
\end{itemize}
The simulation protocol was chosen similar as in ref \cite{FennixBio} (see section V. B). Using the STMV simulations, we open a new window on the impact of pH fluctuations on biological systems, particularly viruses, which is a critical factor influencing their structural dynamics and function. Traditional MD approaches often simulate pH effects by manually adjusting the protonation states of ionizable residues to match the desired pH conditions or rely on specific constant-pH MD implementations\cite{constantphreview} that are strongly dependent on the force field quality. For STMV, as for any other viral particle, pH-dependent changes play a key role in the infection process, likely affecting capsid stability and RNA-capsid interactions. Simulating the pH transition from physiological to more acidic reflects conditions likely to be encountered during the early stages of viral cell entry and intracellular transport\cite{Shen2013}. Implying the "beyond DFT" Foundation Neural Network, we simulated the transition from pH$=$7 to pH$=$5 by explicitly introducing the calculated number of free protons into the system, based on pKa estimations. This approach allows proton exchange to occur dynamically and interactively with the molecular environment. As a proof of concept, we extracted a viral STMV capsomer (33,317 atoms) and quantitatively estimated the number of proton transfer events related to the 6-ns capsomer simulation (Fig. \ref{fig:STMV_capsomer}), providing a dynamic portrait of the molecular response to acidification.

\begin{figure}[htbp] \centering \includegraphics[width=0.45\textwidth]{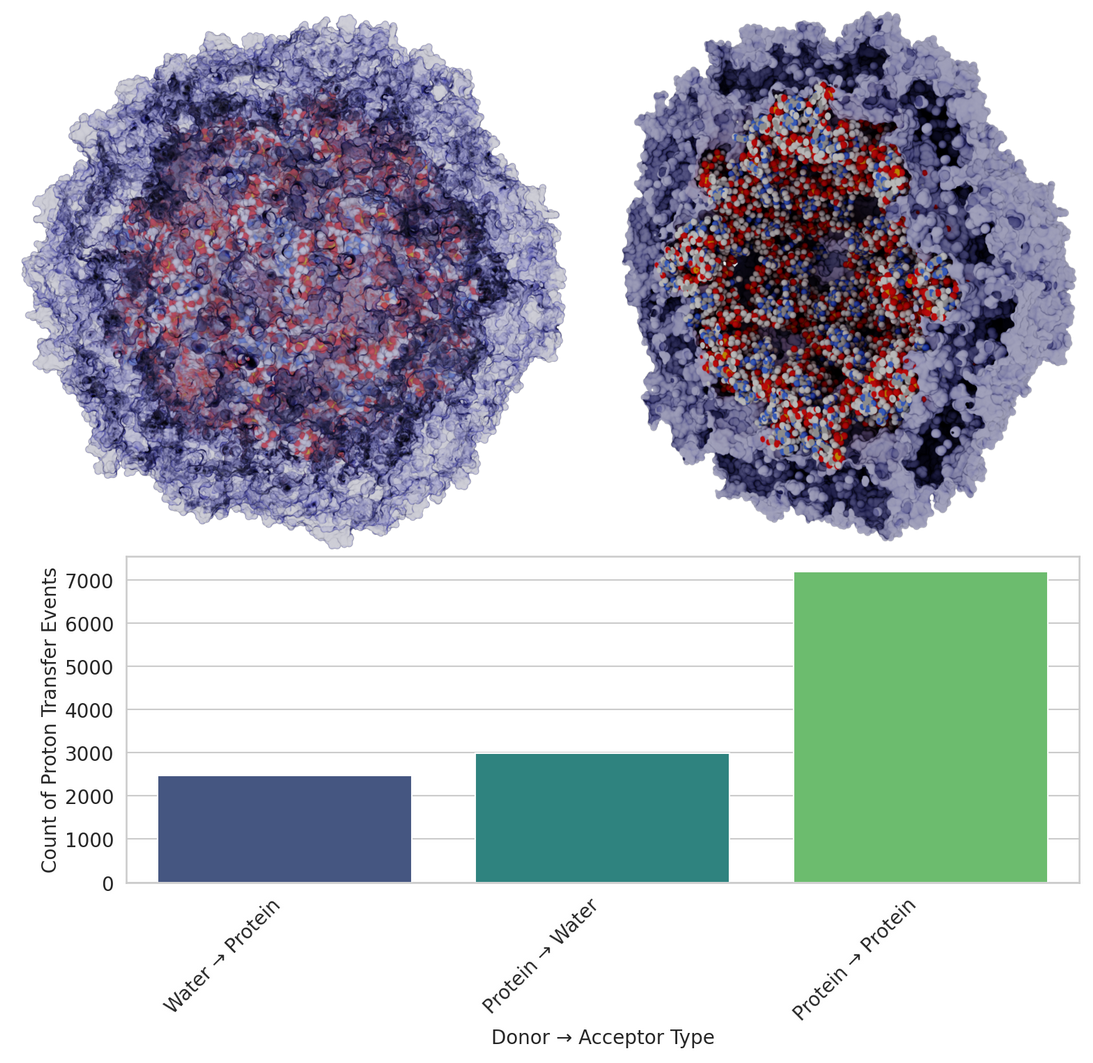} \caption{Number of proton transfer events among key donor–acceptor pairs: water–protein, protein–water, and protein–protein. The protein component corresponds to the STMV capsid. The top panel corresponds to the full STMV nucleocapsid (left) and a cross-sectional view of capsid–RNA assembly  with surrounding ions (right). The nucleocapsid images are generated with the VTX molecular visualization software\cite{Maria2025}.} \label{fig:STMV_full} \end{figure}

Extending this approach to the fully solvated STMV nucleocapsid (of $\sim$ 1 million atoms), we were able to conduct a 1.5-ns on-the-fly viral response to acidification, making a significant step forward in realistic, reactive virus modeling under varying pH conditions. Compared to the isolated capsomer, the current results reveal notable differences (Fig. \ref{fig:STMV_full}). Specifically, we observed an increased number of protein-protein interactions within the capsid, particularly among residues that attempt to relax within the initial frames of the simulation. This behavior underscores the complexity of the full viral system. Additionally, the numerous water molecules present in the system promote frequent proton transfer events within the bulk. Given the large size and intricate nature of STMV, a longer simulation timescale is essential to fully capture its dynamical behavior and is currently underway on the Aurora machine.

\textbf{Perspectives}
We demonstrated the practical applicability of our computational pipeline using FeNNix-Bio1, a new generation foundation NNP model trained with our high-fidelity datasets including QMC/s-CI results. Coupled with RPMD and adaptive sampling, the present implementation will be further optimized allowing to perform million-atom biosystems condensed-phase quantum dynamics simulations approaching the full quantum chemical accuracy. This will enable a closer look at biochemical processes involving chemical reactions. This major step forward bridges the gap between chemically accurate quantum methods and ML simulations for biologically relevant scales. The ML ability to simulate multi-nanosecond quantum MD on such large systems including their chemical reactivity will profoundly impact our understanding of fundamental biochemical processes, potentially revolutionizing fields such as drug discovery, enzyme catalysis, and biomolecular engineering. Future work will explore even larger scales and more diverse chemical spaces, laying the foundation for improved general-purpose ML potentials of unprecedented accuracy and versatility towards predictive molecular modeling in Chemistry, Biology and Drug Design.


\ifCLASSOPTIONcompsoc
  \section*{Acknowledgments}
\else
  \section*{Acknowledgment}
\fi

\footnotesize{
QMC, sCI calculations, DFT computations on Aurora and some MD simulations performed in this research used an award of computer time provided by the Innovative and Novel Computational Impact on Theory and Experiment (INCITE) program (AB, KG, JPP). DFT Calculations on Polaris used  an award of computer time provided by the Innovative and Novel Computational Impact on Theory and Experiment (INCITE) program (RCC, JTK and YL).This research has used resources (Aurora and Polaris machines) of the Argonne Leadership Computing Facility, which is a DOE Office of Science User Facility supported under Contract DE-AC02-06CH11357.  RCC (QMC forces algorithms), JTK (QMC workflow engine) and YL (QMCPACK code development) were supported by the U.S. Department of Energy, Office of Science, Basic Energy Sciences, Materials Sciences and Engineering Division, as part of the Computational Materials Sciences Program and Center for Predictive Simulation of Functional Materials. CK and VA were supported by the Computational Chemical Sciences Program, under Award DE-SC0023382, funded by the U.S. Department of Energy, Office of Basic Energy Sciences, Chemical Sciences, Geosciences, and Biosciences Division.
This work was also made possible thanks to funding from the European Research Council (ERC) under the European Union's Horizon 2020 research and innovation program (grant agreement No 810367), project EMC2 (JPP). We acknowledge EuroHPC Joint Undertaking for awarding the project IDs EHPC-DEV-2024D07-044 (E. P.) and EHPC-AI-2024A04-085 (E. P.) access to Leonardo at CINECA, Italy for DFT computations.  The learning of the FeNNix-Bio1 model and some MD simulations were performed at IDRIS, GENCI (Jean Zay machine, France) on grants No. A0150712052 (J.-P.P.) and grant GC010815453 (Grand Challenge H100 Jean Zay, J.-P.P.). 

This manuscript has been authored by UT-Battelle, LLC under Contract No. DE-AC05-00OR22725 with the U.S. Department of Energy. The United States Government retains and the publisher, by accepting the article for publication, acknowledges that the United States Government retains a non-exclusive, paid-up,irrevocable, worldwide license to publish or reproduce the published form of this manuscript, or allow others to do so, for United States Government purposes. The Department of Energy will provide public access to these results of federally sponsored research in accordance with the DOE Public Access Plan (http://energy.gov/downloads/doe-public-access-plan).
}



%

\bibliographystyle{ieeetr}
\bibliography{GB}

\begin{thebibliography}{100}

\bibitem{Helgaker20082008}
T.~Helgaker, W.~Klopper, and D.~P. Tew, ``Quantitative quantum chemistry,'' {\em Molecular Physics}, vol.~106, no.~16-18, pp.~2107--2143, 2008.

\bibitem{markland2018nuclear}
T.~E. Markland and M.~Ceriotti, ``Nuclear quantum effects enter the mainstream,'' {\em Nature Reviews Chemistry}, vol.~2, no.~3, p.~0109, 2018.

\bibitem{yang2024machine}
Y.~Yang, S.~Zhang, K.~D. Ranasinghe, O.~Isayev, and A.~E. Roitberg, ``Machine learning of reactive potentials,'' {\em Annual Review of Physical Chemistry}, vol.~75, no.~1, pp.~371--395, 2024.

\bibitem{marx2009ab}
D.~Marx and J.~Hutter, {\em Ab initio molecular dynamics: basic theory and advanced methods}.
\newblock Cambridge University Press, 2009.

\bibitem{MACKERELL200591}
A.~D. MacKerell, ``Chapter 7 empirical force fields for proteins: Current status and future directions,'' 2005.

\bibitem{jia2020pushing}
W.~Jia, H.~Wang, M.~Chen, D.~Lu, L.~Lin, R.~Car, E.~Weinan, and L.~Zhang, ``Pushing the limit of molecular dynamics with ab initio accuracy to 100 million atoms with machine learning,'' in {\em SC20: International conference for high performance computing, networking, storage and analysis}, pp.~1--14, IEEE, 2020.

\bibitem{batatia2024foundationmodelatomisticmaterials}
I.~Batatia, P.~Benner, Y.~Chiang, A.~M. Elena, D.~P. Kov{\'a}cs, J.~Riebesell, X.~R. Advincula, M.~Asta, M.~Avaylon, W.~J. Baldwin, {\em et~al.}, ``A foundation model for atomistic materials chemistry,'' 2024.
\newblock preprint ArXiv 2401.00096.

\bibitem{mauger2021nuclear}
N.~Mauger, T.~Pl{\'e}, L.~Lagard{\`e}re, S.~Bonella, {\'E}.~Mangaud, J.-P. Piquemal, and S.~Huppert, ``Nuclear quantum effects in liquid water at near classical computational cost using the adaptive quantum thermal bath,'' {\em The Journal of Physical Chemistry Letters}, vol.~12, no.~34, pp.~8285--8291, 2021.

\bibitem{RAGHAVACHARI1989479}
K.~Raghavachari, G.~W. Trucks, J.~A. Pople, and M.~Head-Gordon, ``A fifth-order perturbation comparison of electron correlation theories,'' {\em Chemical Physics Letters}, vol.~157, no.~6, pp.~479--483, 1989.

\bibitem{schafer2024}
T.~Schäfer, A.~Irmler, A.~Gallo, and A.~Grüneis, ``Understanding discrepancies of wavefunction theories for large molecules,'' 2024.
\newblock Preprint arXiv:2407.01442.

\bibitem{legeza2025}
Örs Legeza, A.~Menczer, Ádám Ganyecz, M.~A. Werner, K.~Kapás, J.~Hammond, S.~S. Xantheas, M.~Ganahl, and F.~Neese, ``Orbital optimization of large active spaces via ai-accelerators,'' 2025.
\newblock preprint arXiv:2503.20700.

\bibitem{Reynolds_1982}
P.~J. Reynolds, D.~M. Ceperley, B.~J. Alder, and W.~A. Lester, ``Fixed‐node quantum monte carlo for molecules,'' {\em J. Chem. Phys.}, vol.~77, pp.~5593--5603, 1982.

\bibitem{Reynolds1990}
P.~J. Reynolds, J.~Tobochnik, and H.~Gould, ``Diffusion quantum monte carlo,'' {\em Computers in Physics}, vol.~4, no.~6, pp.~662--668, 1990.

\bibitem{foulkes01}
W.~M.~C. Foulkes, L.~Mitas, R.~J. Needs, and G.~Rajagopal, ``Quantum monte carlo simulations of solids,'' {\em Rev. Mod. Phys.}, vol.~73, pp.~33--83, 2001.

\bibitem{foyevtsova14}
K.~Foyevtsova, J.~T. Krogel, J.~Kim, P.~R.~C. Kent, E.~Dagotto, and F.~A. Reboredo, ``Ab initio quantum monte carlo calculations of spin superexchange in cuprates: The benchmarking case of ca$_2$cuo$_3$,'' {\em Phys. Rev. X}, vol.~4, p.~031003, 2014.

\bibitem{shin17}
H.~Shin, Y.~Luo, P.~Ganesh, J.~Balachandran, J.~T. Krogel, P.~R.~C. Kent, A.~Benali, and O.~Heinonen, ``Electronic properties of doped and defective nio: A quantum monte carlo study,'' {\em Phys.\ Rev.\ Materials}, vol.~1, p.~173603, 2017.

\bibitem{shin2017}
H.~Shin, J.~Kim, H.~Lee, O.~Heinonen, A.~Benali, and Y.~Kwon, ``Nature of interlayer binding and stacking of sp$-$sp$^2$ hybridized carbon layers: A quantum monte carlo study,'' {\em J. Chem. Theory Comput.}, vol.~13, no.~11, pp.~5639--5646, 2017.

\bibitem{kylanpaa17}
I.~Kyl$\text{\"{a}}$np$\text{\"{a}\"{a}}$, J.~Balachandran, G.~Panchapakesan, O.~Heinonen, P.~R.~C. Kent, and J.~T. Krogel, ``Accuracy of ab initio electron correlation and electron densities in vanadium dioxide,'' {\em Phys.\ Rev.\ Materials}, vol.~1, p.~065408, 2017.

\bibitem{shin18}
H.~Shin, A.~Benali, Y.~Luo, E.~Crabb, A.~Lopez-Bezanilla, L.~E. Ratcliff, A.~M. Jokisaari, and O.~Heinonen, ``Zirconia and hafnia polymorphs: Ground-state structural properties from diffusion monte carlo,'' {\em Phys.\ Rev.\ Materials}, vol.~2, p.~075001, 2018.

\bibitem{Morales2012}
M.~A. Morales, J.~McMinis, B.~K. Clark, J.~Kim, and G.~E. Scuseria, ``Multideterminant wave functions in quantum monte carlo,'' {\em Journal of Chemical Theory and Computation}, vol.~8, no.~7, pp.~2181--2188, 2012.

\bibitem{Benali_2014}
A.~Benali, L.~Shulenburger, N.~A. Romero, J.~Kim, and O.~A. von Lilienfeld, ``Application of diffusion monte carlo to materials dominated by van der waals interactions,'' {\em J.\ Chem.\ Theory\ Comput.}, vol.~10, no.~8, pp.~3417--3422, 2014.

\bibitem{benali16}
A.~Benali, L.~Shulenburger, J.~T. Krogel, X.~Zhong, P.~R.~C. Kent, and O.~Heinonen, ``Quantum monte carlo analysis of a charge ordered insulating antiferromagnet: the ti$_4$o$_7$ magn\'{e}li phase,'' {\em Phys.\ Chem.\ Chem.\ Phys.}, vol.~18, pp.~18323--18335, 2016.

\bibitem{Kolorenvc_2011}
J.~Koloren{\v{c}} and L.~Mitas, ``Applications of quantum monte carlo methods in condensed systems,'' {\em Reports on Progress in Physics}, vol.~74, no.~2, p.~026502, 2011.

\bibitem{Dubecky_2016}
M.~Dubeck{\'y}, L.~Mitas, and P.~Jure{\v c}ka, ``Noncovalent {Interactions} by {Quantum} {Monte} {Carlo},'' {\em Chemical Reviews}, vol.~116, pp.~5188--5215, May 2016.

\bibitem{Benali2020}
A.~Benali, K.~Gasperich, K.~D. Jordan, T.~Applencourt, Y.~Luo, M.~C. Bennett, J.~T. Krogel, L.~Shulenburger, P.~R.~C. Kent, P.-F. Loos, A.~Scemama, and M.~Caffarel, ``Toward a systematic improvement of the fixed-node approximation in diffusion monte carlo for solids—a case study in diamond,'' {\em The Journal of Chemical Physics}, vol.~153, no.~18, p.~184111, 2020.

\bibitem{khan2024}
D.~Khan, A.~Benali, S.~Y.~H. Kim, G.~F. von Rudorff, and O.~A. von Lilienfeld, ``Quantum mechanical dataset of 836k neutral closed shell molecules with upto 5 heavy atoms from cnofsipsclbr,'' 2024.

\bibitem{Filippi2000}
C.~Filippi and C.~J. Umrigar, ``Correlated sampling in quantum monte carlo: A route to forces,'' {\em Phys. Rev. B}, vol.~61, pp.~R16291--R16294, Jun 2000.

\bibitem{Wagner2014}
L.~K. Wagner and D.~Pathak, ``Improved quantum {M}onte {C}arlo forces for interaction energy curves and gap energetics,'' {\em The Journal of Chemical Physics}, vol.~141, no.~6, p.~064102, 2014.

\bibitem{Filippi2016}
C.~Filippi, R.~Assaraf, and S.~Moroni, ``Simple formalism for efficient derivatives and multi-determinant expansions in quantum monte carlo,'' {\em The Journal of Chemical Physics}, vol.~144, no.~19, p.~194105, 2016.

\bibitem{Caffarelforces2000}
R.~Assaraf and M.~Caffarel, ``Computing forces with quantum monte carlo,'' {\em The Journal of Chemical Physics}, vol.~113, pp.~4028--4034, 09 2000.

\bibitem{Caffarel2013}
E.~Giner, A.~Scemama, and M.~Caffarel, ``Using perturbatively selected configuration interaction in quantum monte carlo calculations,'' {\em Canadian Journal of Chemistry}, vol.~91, no.~9, pp.~879--885, 2013.

\bibitem{HURON1974277}
B.~Huron, J.~Malrieu, and P.~Rancurel, ``Perturbation calculation of transition moments. application to h2 and mgo,'' {\em Chemical Physics}, vol.~3, no.~2, pp.~277--283, 1974.

\bibitem{Tubman2016}
N.~M. Tubman, J.~Lee, T.~Y. Takeshita, M.~Head-Gordon, and K.~B. Whaley, ``A deterministic alternative to the full configuration interaction quantum monte carlo method,'' {\em The Journal of Chemical Physics}, vol.~145, no.~4, p.~044112, 2016.

\bibitem{Holmes_2017}
A.~A. Holmes, C.~J. Umrigar, and S.~Sharma, ``Excited states using semistochastic heat-bath configuration interaction,'' {\em J. Chem. Phys.}, vol.~147, p.~164111, Oct 2017.

\bibitem{Scemama_2018b}
A.~Scemama, A.~Benali, D.~Jacquemin, M.~Caffarel, and P.-F. Loos, ``Excitation energies from diffusion monte carlo using selected configuration interaction nodes,'' {\em J. Chem. Phys.}, vol.~149, p.~034108, jul 2018.

\bibitem{Loos_2018b}
P.-F. Loos, A.~Scemama, A.~Blondel, Y.~Garniron, M.~Caffarel, and D.~Jacquemin, ``A mountaineering strategy to excited states: Highly accurate reference energies and benchmarks,'' {\em J. Chem. Theory Comput.}, vol.~14, pp.~4360--4379, jul 2018.

\bibitem{Loos_2020b}
P.~F. Loos, F.~Lipparini, M.~Boggio-Pasqua, A.~Scemama, and D.~Jacquemin, ``A mountaineering strategy to excited states: highly-accurate energies and benchmarks for medium size molecules,,'' {\em J. Chem. Theory Comput.}, vol.~16, p.~1711, 2020.

\bibitem{Loos_2020a}
P.-F. Loos, A.~Scemama, and D.~Jacquemin, ``The quest for highly accurate excitation energies: A computational perspective,'' {\em The Journal of Physical Chemistry Letters}, vol.~11, no.~6, pp.~2374--2383, 2020.
\newblock PMID: 32125872.

\bibitem{Scemama_2019}
A.~Scemama, M.~Caffarel, A.~Benali, D.~Jacquemin, and P.-F. Loos, ``{Influence of pseudopotentials on excitation energies from selected configuration interaction and diffusion Monte Carlo},'' {\em Res. Chem.}, vol.~1, p.~100002, 2019.

\bibitem{Loos_2019}
P.-F. Loos, M.~Boggio-Pasqua, A.~Scemama, M.~Caffarel, and D.~Jacquemin, ``Reference energies for double excitations,'' {\em Journal of Chemical Theory and Computation}, vol.~15, no.~3, pp.~1939--1956, 2019.

\bibitem{Neuscamman2020}
L.~Otis, I.~Craig, and E.~Neuscamman, ``A hybrid approach to excited-state-specific variational monte carlo and doubly excited states,'' {\em The Journal of Chemical Physics}, vol.~153, no.~23, p.~234105, 2020.

\bibitem{neuscamman2019}
S.~D. Pineda~Flores and E.~Neuscamman, ``Excited state specific multi-slater jastrow wave functions,'' {\em The Journal of Physical Chemistry A}, vol.~123, no.~8, pp.~1487--1497, 2019.
\newblock PMID: 30702890.

\bibitem{Garner2020}
S.~M. Garner and E.~Neuscamman, ``A variational monte carlo approach for core excitations,'' {\em The Journal of Chemical Physics}, vol.~153, no.~14, p.~144108, 2020.

\bibitem{Slootman2024}
E.~Slootman, I.~Poltavsky, R.~Shinde, J.~Cocomello, S.~Moroni, A.~Tkatchenko, and C.~Filippi, ``Accurate quantum monte carlo forces for machine-learned force fields: Ethanol as a benchmark,'' {\em Journal of Chemical Theory and Computation}, vol.~20, no.~14, pp.~6020--6027, 2024.

\bibitem{QMCPACK1}
J.~Kim, A.~D. Baczewski, T.~D. Beaudet, A.~Benali, M.~C. Bennett, M.~A. Berrill, N.~S. Blunt, E.~J.~L. Borda, M.~Casula, D.~M. Ceperley, {\em et~al.}, ``Qmcpack : An open source ab initio quantum monte carlo package for the electronic structure of atoms, molecules, and solids,'' {\em J. Phys.: Condens. Matter}, vol.~30, p.~195901, 2018.

\bibitem{QMCPACK2}
P.~R.~C. Kent, A.~Annaberdiyev, A.~Benali, M.~C. Bennett, E.~J.~L. Borda, P.~Doak, H.~Hao, K.~D. Jordan, J.~T. Krogel, I.~Kyl{\"{a}}np{\"{a}}{\"{a}}, J.~Lee, Y.~Luo, F.~D. Malone, C.~A. Melton, L.~Mitas, M.~A. Morales, E.~Neuscamman, F.~A. Reboredo, B.~Rubenstein, K.~Saritas, S.~Upadhyay, G.~Wang, S.~Zhang, and L.~Zhao, ``{QMCPACK}: Advances in the development, efficiency, and application of auxiliary field and real-space variational and diffusion quantum monte carlo,'' {\em The Journal of Chemical Physics}, vol.~152, p.~174105, May 2020.

\bibitem{QMCPACK3}
A.~Mathuriya, Y.~Luo, R.~C. Clay, A.~Benali, L.~Shulenburger, and J.~Kim, ``Embracing a new era of highly efficient and productive quantum monte carlo simulations,'' in {\em Proceedings of the International Conference for High Performance Computing, Networking, Storage and Analysis}, SC '17, (New York, NY, USA), Association for Computing Machinery, 2017.

\bibitem{QMCPACK4}
A.~Mathuriya, Y.~Luo, A.~Benali, L.~Shulenburger, and J.~Kim, ``Optimization and parallelization of b-spline based orbital evaluations in qmc on multi/many-core shared memory processors,'' in {\em 2017 IEEE International Parallel and Distributed Processing Symposium (IPDPS)}, pp.~213--223, 2017.

\bibitem{behler2007generalized}
J.~Behler and M.~Parrinello, ``Generalized neural-network representation of high-dimensional potential-energy surfaces,'' {\em Physical Review Letters}, vol.~98, no.~14, p.~146401, 2007.

\bibitem{bartok2010gaussian}
A.~P. Bart{\'o}k, M.~C. Payne, R.~Kondor, and G.~Cs{\'a}nyi, ``Gaussian approximation potentials: The accuracy of quantum mechanics, without the electrons,'' {\em Physical Review Letters}, vol.~104, no.~13, p.~136403, 2010.

\bibitem{chmiela2018towards}
S.~Chmiela, H.~E. Sauceda, K.-R. M{\"u}ller, and A.~Tkatchenko, ``Towards exact molecular dynamics simulations with machine-learned force fields,'' {\em Nature Communications}, vol.~9, no.~1, pp.~1--10, 2018.

\bibitem{bigi2023wigner}
F.~Bigi, S.~N. Pozdnyakov, and M.~Ceriotti, ``Wigner kernels: body-ordered equivariant machine learning without a basis,'' {\em arXiv preprint arXiv:2303.04124}, 2023.

\bibitem{shakouri2017accurate}
K.~Shakouri, J.~Behler, J.~Meyer, and G.-J. Kroes, ``Accurate neural network description of surface phonons in reactive gas--surface dynamics: N2+ ru (0001),'' {\em The journal of physical chemistry letters}, vol.~8, no.~10, pp.~2131--2136, 2017.

\bibitem{smith2017ani}
J.~S. Smith, O.~Isayev, and A.~E. Roitberg, ``Ani-1: an extensible neural network potential with dft accuracy at force field computational cost,'' {\em Chemical Science}, vol.~8, no.~4, pp.~3192--3203, 2017.

\bibitem{schutt2017quantum}
K.~T. Sch{\"u}tt, F.~Arbabzadah, S.~Chmiela, K.~R. M{\"u}ller, and A.~Tkatchenko, ``Quantum-chemical insights from deep tensor neural networks,'' {\em Nature communications}, vol.~8, no.~1, p.~13890, 2017.

\bibitem{schutt2017schnet}
K.~Sch{\"u}tt, P.-J. Kindermans, H.~E. Sauceda~Felix, S.~Chmiela, A.~Tkatchenko, and K.-R. M{\"u}ller, ``Schnet: A continuous-filter convolutional neural network for modeling quantum interactions,'' {\em Advances in neural information processing systems}, vol.~30, 2017.

\bibitem{gilmer2017neural}
J.~Gilmer, S.~S. Schoenholz, P.~F. Riley, O.~Vinyals, and G.~E. Dahl, ``Neural message passing for quantum chemistry,'' in {\em International conference on machine learning}, pp.~1263--1272, PMLR, 2017.

\bibitem{lubbers2018hierarchical}
N.~Lubbers, J.~S. Smith, and K.~Barros, ``Hierarchical modeling of molecular energies using a deep neural network,'' {\em The Journal of Chemical Physics}, vol.~148, no.~24, 2018.

\bibitem{zubatyuk2019accurate}
R.~Zubatyuk, J.~S. Smith, J.~Leszczynski, and O.~Isayev, ``Accurate and transferable multitask prediction of chemical properties with an atoms-in-molecules neural network,'' {\em Science advances}, vol.~5, no.~8, p.~eaav6490, 2019.

\bibitem{zaverkin2020gaussian}
V.~Zaverkin and J.~K{\"a}stner, ``Gaussian moments as physically inspired molecular descriptors for accurate and scalable machine learning potentials,'' {\em Journal of Chemical Theory and Computation}, vol.~16, no.~8, pp.~5410--5421, 2020.

\bibitem{qiao2021unite}
Z.~Qiao, A.~S. Christensen, M.~Welborn, F.~R. Manby, A.~Anandkumar, and T.~F. Miller~III, ``Unite: Unitary n-body tensor equivariant network with applications to quantum chemistry,'' {\em arXiv preprint arXiv:2105.14655}, 2021.

\bibitem{unke2021spookynet}
O.~T. Unke, S.~Chmiela, M.~Gastegger, K.~T. Sch{\"u}tt, H.~E. Sauceda, and K.-R. M{\"u}ller, ``Spookynet: Learning force fields with electronic degrees of freedom and nonlocal effects,'' {\em Nature Communications}, vol.~12, no.~1, pp.~1--14, 2021.

\bibitem{gasteiger2021gemnet}
J.~Gasteiger, F.~Becker, and S.~G{\"u}nnemann, ``Gemnet: Universal directional graph neural networks for molecules,'' {\em Advances in Neural Information Processing Systems}, vol.~34, pp.~6790--6802, 2021.

\bibitem{batzner20223}
S.~Batzner, A.~Musaelian, L.~Sun, M.~Geiger, J.~P. Mailoa, M.~Kornbluth, N.~Molinari, T.~E. Smidt, and B.~Kozinsky, ``E(3)-equivariant graph neural networks for data-efficient and accurate interatomic potentials,'' {\em Nature Communications}, vol.~13, no.~1, pp.~1--11, 2022.

\bibitem{musaelian2022learning}
A.~Musaelian, S.~Batzner, A.~Johansson, L.~Sun, C.~J. Owen, M.~Kornbluth, and B.~Kozinsky, ``Learning local equivariant representations for large-scale atomistic dynamics,'' {\em Nature Communications}, vol.~14, no.~1, p.~579, 2023.

\bibitem{grisafi2019incorporating}
A.~Grisafi and M.~Ceriotti, ``Incorporating long-range physics in atomic-scale machine learning,'' {\em The Journal of Chemical Physics}, vol.~151, no.~20, p.~204105, 2019.

\bibitem{grisafi2021multi}
A.~Grisafi, J.~Nigam, and M.~Ceriotti, ``Multi-scale approach for the prediction of atomic scale properties,'' {\em Chemical Science}, vol.~12, no.~6, pp.~2078--2090, 2021.

\bibitem{ANI2X}
C.~Devereux, J.~S. Smith, K.~K. Huddleston, K.~Barros, R.~Zubatyuk, O.~Isayev, and A.~E. Roitberg, ``Extending the applicability of the ani deep learning molecular potential to sulfur and halogens,'' {\em Journal of Chemical Theory and Computation}, vol.~16, no.~7, pp.~4192--4202, 2020.

\bibitem{AIMNET}
R.~Zubatyuk, J.~S. Smith, J.~Leszczynski, and O.~Isayev, ``Accurate and transferable multitask prediction of chemical properties with an atoms-in-molecules neural network,'' {\em Science Advances}, vol.~5, no.~8, p.~eaav6490, 2019.

\bibitem{unke2024biomolecular}
O.~T. Unke, M.~St{\"o}hr, S.~Ganscha, T.~Unterthiner, H.~Maennel, S.~Kashubin, D.~Ahlin, M.~Gastegger, L.~Medrano~Sandonas, J.~T. Berryman, {\em et~al.}, ``Biomolecular dynamics with machine-learned quantum-mechanical force fields trained on diverse chemical fragments,'' {\em Science Advances}, vol.~10, no.~14, p.~eadn4397, 2024.

\bibitem{pinheiro2021choosing}
M.~Pinheiro, F.~Ge, N.~Ferr{\'e}, P.~O. Dral, and M.~Barbatti, ``Choosing the right molecular machine learning potential,'' {\em Chemical Science}, vol.~12, no.~43, pp.~14396--14413, 2021.

\bibitem{kocer2022neural}
E.~Kocer, T.~W. Ko, and J.~Behler, ``Neural network potentials: A concise overview of methods,'' {\em Annual review of physical chemistry}, vol.~73, no.~1, pp.~163--186, 2022.

\bibitem{lammpsnn}
A.~Singraber, J.~Behler, and C.~Dellago, ``Library-based lammps implementation of high-dimensional neural network potentials,'' {\em Journal of Chemical Theory and Computation}, vol.~15, no.~3, pp.~1827--1840, 2019.
\newblock PMID: 30677296.

\bibitem{inizan2022scalable}
T.~Jaffrelot~Inizan, T.~Pl\'e, O.~Adjoua, P.~Ren, H.~Gokcan, O.~Isayev, L.~Lagard\`ere, and J.-P. Piquemal, ``Scalable hybrid deep neural networks/polarizable potentials biomolecular simulations including long-range effects,'' {\em Chemical Science}, vol.~14, pp.~5438--5452, 2023.

\bibitem{schutt2023schnetpack}
K.~T. Sch{\"u}tt, S.~S. Hessmann, N.~W. Gebauer, J.~Lederer, and M.~Gastegger, ``Schnetpack 2.0: A neural network toolbox for atomistic machine learning,'' {\em The Journal of Chemical Physics}, vol.~158, no.~14, 2023.

\bibitem{dral2024mlatom}
P.~O. Dral, F.~Ge, Y.-F. Hou, P.~Zheng, Y.~Chen, M.~Barbatti, O.~Isayev, C.~Wang, B.-X. Xue, M.~Pinheiro~Jr, {\em et~al.}, ``Mlatom 3: A platform for machine learning-enhanced computational chemistry simulations and workflows,'' {\em Journal of Chemical Theory and Computation}, vol.~20, no.~3, pp.~1193--1213, 2024.

\bibitem{zeng2023deepmd}
J.~Zeng, D.~Zhang, D.~Lu, P.~Mo, Z.~Li, Y.~Chen, M.~Rynik, L.~Huang, Z.~Li, S.~Shi, {\em et~al.}, ``Deepmd-kit v2: A software package for deep potential models,'' {\em The Journal of Chemical Physics}, vol.~159, no.~5, 2023.

\bibitem{ple2024fennol}
T.~Pl\'e, O.~Adjoua, L.~Lagard\`ere, and J.-P. Piquemal, ``{FeNNol: An efficient and flexible library for building force-field-enhanced neural network potentials},'' {\em The Journal of Chemical Physics}, vol.~161, p.~042502, 07 2024.

\bibitem{alphafold}
J.~Jumper, R.~Evans, A.~Pritzel, T.~Green, M.~Figurnov, O.~Ronneberger, K.~Tunyasuvunakool, R.~Bates, A.~{\v{Z}}{\'\i}dek, A.~Potapenko, {\em et~al.}, ``Highly accurate protein structure prediction with alphafold,'' {\em Nature}, vol.~596, no.~7873, pp.~583--589, 2021.

\bibitem{rosettafold}
M.~Baek, F.~DiMaio, I.~Anishchenko, J.~Dauparas, S.~Ovchinnikov, G.~R. Lee, J.~Wang, Q.~Cong, L.~N. Kinch, R.~D. Schaeffer, {\em et~al.}, ``Accurate prediction of protein structures and interactions using a three-track neural network,'' {\em Science}, vol.~373, no.~6557, pp.~871--876, 2021.

\bibitem{FennixBio}
T.~Plé, O.~Adjoua, A.~Benali, E.~Posenitskiy, C.~Villot, L.~Lagardère, and J.-P. Piquemal, ``A foundation model for accurate atomistic simulations in drug design,'' {\em ChemRxiv}, 2025.
\newblock DOI: 10.26434/chemrxiv-2025-f1hgn.

\bibitem{D3SC02581K}
T.~Plé, L.~Lagardère, and J.-P. Piquemal, ``Force-field-enhanced neural network interactions: from local equivariant embedding to atom-in-molecule properties and long-range effects,'' {\em Chem. Sci.}, vol.~14, pp.~12554--12569, 2023.

\bibitem{ceriotti2016nuclear}
M.~Ceriotti, W.~Fang, P.~G. Kusalik, R.~H. McKenzie, A.~Michaelides, M.~A. Morales, and T.~E. Markland, ``Nuclear quantum effects in water and aqueous systems: Experiment, theory, and current challenges,'' {\em Chemical Reviews}, vol.~116, no.~13, pp.~7529--7550, 2016.

\bibitem{ple2023routine}
T.~Pl{\'e}, N.~Mauger, O.~Adjoua, T.~J. Inizan, L.~Lagard{\`e}re, S.~Huppert, and J.-P. Piquemal, ``Routine molecular dynamics simulations including nuclear quantum effects: from force fields to machine learning potentials,'' {\em Journal of Chemical Theory and Computation}, vol.~19, no.~5, pp.~1432--1445, 2023.

\bibitem{RevModPhys.71.1253}
W.~Kohn, ``Nobel lecture: Electronic structure of matter---wave functions and density functionals,'' {\em Rev. Mod. Phys.}, vol.~71, pp.~1253--1266, Oct 1999.

\bibitem{FCIlimit}
K.~D. Vogiatzis, D.~Ma, J.~Olsen, L.~Gagliardi, and W.~A. de~Jong, ``Pushing configuration-interaction to the limit: Towards massively parallel mcscf calculations,'' {\em The Journal of Chemical Physics}, vol.~147, p.~184111, 11 2017.

\bibitem{Spice2_1}
P.~Eastman, P.~K. Behara, D.~L. Dotson, Z.~Meng, C.~Velez-Vega, S.~Zhang, A.~Jain, C.~Kramer, M.~J. Robertson, J.~Swails, J.~D. Chodera, B.~Sheets, R.~O. Drorr, and V.~S. Pande, ``Spice, a dataset of drug-like molecules and peptides for training machine learning potentials,'' {\em Scientific Data}, vol.~10, no.~1, p.~193, 2023.

\bibitem{eastman2023spice}
P.~Eastman, P.~K. Behara, D.~L. Dotson, R.~Galvelis, J.~E. Herr, J.~T. Horton, Y.~Mao, J.~D. Chodera, B.~P. Pritchard, Y.~Wang, {\em et~al.}, ``Spice, a dataset of drug-like molecules and peptides for training machine learning potentials,'' {\em Scientific Data}, vol.~10, no.~1, p.~11, 2023.

\bibitem{ccECP}
G.~Wang, A.~Annaberdiyev, C.~A. Melton, M.~C. Bennett, L.~Shulenburger, and L.~Mitas, ``A new generation of effective core potentials from correlated calculations: 4s and 4p main group elements and first row additions,'' {\em The Journal of Chemical Physics}, vol.~151, p.~144110, 10 2019.

\bibitem{PySCF}
Q.~Sun, T.~C. Berkelbach, N.~S. Blunt, G.~H. Booth, S.~Guo, Z.~Li, J.~Liu, J.~D. McClain, E.~R. Sayfutyarova, S.~Sharma, S.~Wouters, and G.~K. Chan, ``{{PySCF: the Python‐Based Simulations of Chemistry Framework}},'' {\em WIREs Comput. Mol. Sci.}, vol.~8, no.~1, p.~e1340, 2018.

\bibitem{pyscf_cpu}
Q.~Sun, T.~C. Berkelbach, N.~S. Blunt, G.~H. Booth, S.~Guo, Z.~Li, J.~Liu, J.~D. McClain, E.~R. Sayfutyarova, S.~Sharma, S.~Wouters, and G.~K. Chan, ``Pyscf: the python‐based simulations of chemistry framework,'' 2017.

\bibitem{Pyscf_gpu1}
R.~Li, Q.~Sun, X.~Zhang, and G.~K.-L. Chan, ``Introducing gpu acceleration into the python-based simulations of chemistry framework,'' {\em The Journal of Physical Chemistry A}, vol.~129, no.~5, pp.~1459--1468, 2025.
\newblock PMID: 39846468.

\bibitem{Pyscf_gpu2}
X.~Wu, Q.~Sun, Z.~Pu, T.~Zheng, W.~Ma, W.~Yan, Y.~Xia, Z.~Wu, M.~Huo, X.~Li, W.~Ren, S.~Gong, Y.~Zhang, and W.~Gao, ``Enhancing gpu-acceleration in the python-based simulations of chemistry frameworks,'' {\em WIREs Computational Molecular Science}, vol.~15, no.~2, p.~e70008, 2025.
\newblock e70008 CMS-1146.R2.

\bibitem{Mardirossian2016}
N.~Mardirossian and M.~Head-Gordon, ``$\omega$b97m-v: A combinatorially optimized, range-separated hybrid, meta-gga density functional with vv10 nonlocal correlation,'' {\em The Journal of Chemical Physics}, vol.~144, no.~21, p.~214110, 2016.

\bibitem{Mardirossian2017}
N.~Mardirossian and M.~Head-Gordon, ``Thirty years of density functional theory in computational chemistry: an overview and extensive assessment of 200 density functionals,'' {\em Molecular Physics}, vol.~115, no.~19, pp.~2315--2372, 2017.

\bibitem{Grimme2011}
S.~Grimme, S.~Ehrlich, and L.~Goerigk, ``Effect of the damping function in dispersion corrected density functional theory,'' {\em Journal of Computational Chemistry}, vol.~32, no.~7, pp.~1456--1465, 2011.

\bibitem{Bennett2017}
M.~C. Bennett, C.~Wang, A.~Annaberdiyev, C.~A. Melton, L.~Shulenburger, and L.~Mitas, ``A new generation of effective core potentials from correlated calculations: 2nd row elements,'' {\em The Journal of Chemical Physics}, vol.~147, no.~22, p.~224106, 2017.

\bibitem{Annaberdiyev2018}
A.~Annaberdiyev, G.~Wang, C.~A. Melton, M.~C. Bennett, L.~Shulenburger, and L.~Mitas, ``A new generation of effective core potentials from correlated calculations: 3d transition metal series,'' {\em The Journal of Chemical Physics}, vol.~149, no.~13, p.~134108, 2018.

\bibitem{Anderson1980}
J.~B. Anderson, ``Quantum chemistry by random walk: Higher accuracy,'' {\em The Journal of Chemical Physics}, vol.~73, no.~8, pp.~3897--3899, 1980.

\bibitem{krogel2016}
J.~T. Krogel, ``Nexus: A modular workflow management system for quantum simulation codes,'' {\em Computer Physics Communications}, vol.~198, pp.~154 -- 168, 2016.

\bibitem{zhang2025fault}
Y.~Zhang, X.~Zhang, J.~Sun, H.~Lin, Y.~Huang, D.~Lv, and X.~Yuan, ``Fault-tolerant quantum algorithms for quantum molecular systems: A survey,'' {\em arXiv preprint arXiv:2502.02139}, 2025.

\bibitem{lee2023evaluating}
S.~Lee, J.~Lee, H.~Zhai, Y.~Tong, A.~M. Dalzell, A.~Kumar, P.~Helms, J.~Gray, Z.-H. Cui, W.~Liu, {\em et~al.}, ``Evaluating the evidence for exponential quantum advantage in ground-state quantum chemistry,'' {\em Nature communications}, vol.~14, no.~1, p.~1952, 2023.

\bibitem{traore2024shortcut}
D.~Traore, O.~Adjoua, C.~Feniou, I.-M. Lygatsika, Y.~Maday, E.~Posenitskiy, K.~Hammernik, A.~Peruzzo, J.~Toulouse, E.~Giner, {\em et~al.}, ``Shortcut to chemically accurate quantum computing via density-based basis-set correction,'' {\em Communications Chemistry}, vol.~7, no.~1, p.~269, 2024.

\bibitem{lagardere2018tinker}
L.~Lagard\`ere, L.-H. Jolly, F.~Lipparini, F.~Aviat, B.~Stamm, Z.~F. Jing, M.~Harger, H.~Torabifard, G.~A. Cisneros, M.~J. Schnieders, N.~Gresh, Y.~Maday, P.~Y. Ren, J.~W. Ponder, and J.-P. Piquemal, ``Tinker-hp: a massively parallel molecular dynamics package for multiscale simulations of large complex systems with advanced point dipole polarizable force fields,'' {\em Chemical Science}, vol.~9, pp.~956--972, 2018.

\bibitem{adjoua2021tinker}
O.~Adjoua, L.~Lagard{\`e}re, L.-H. Jolly, A.~Durocher, T.~Very, I.~Dupays, Z.~Wang, T.~Jaffrelot~Inizan, F.~C{\'e}lerse, P.~Ren, J.~W. Ponder, and J.-P. Piquemal, ``Tinker-hp: Accelerating molecular dynamics simulations of large complex systems with advanced point dipole polarizable force fields using gpus and multi-gpu systems,'' {\em Journal of Chemical Theory and Computation}, vol.~17, no.~4, pp.~2034--2053, 2021.

\bibitem{smith2019approaching}
J.~S. Smith, B.~T. Nebgen, R.~Zubatyuk, N.~Lubbers, C.~Devereux, K.~Barros, S.~Tretiak, O.~Isayev, and A.~E. Roitberg, ``Approaching coupled cluster accuracy with a general-purpose neural network potential through transfer learning,'' {\em Nature communications}, vol.~10, no.~1, p.~2903, 2019.

\bibitem{zhuang2020adabelief}
J.~Zhuang, T.~Tang, Y.~Ding, S.~C. Tatikonda, N.~Dvornek, X.~Papademetris, and J.~Duncan, ``Adabelief optimizer: Adapting stepsizes by the belief in observed gradients,'' {\em Advances in neural information processing systems}, vol.~33, pp.~18795--18806, 2020.

\bibitem{tschumper2002anchoring}
G.~S. Tschumper, M.~L. Leininger, B.~C. Hoffman, E.~F. Valeev, H.~F. Schaefer~III, and M.~Quack, ``Anchoring the water dimer potential energy surface with explicitly correlated computations and focal point analyses,'' {\em The Journal of chemical physics}, vol.~116, no.~2, pp.~690--701, 2002.

\bibitem{bates2009ccsd}
D.~M. Bates and G.~S. Tschumper, ``Ccsd (t) complete basis set limit relative energies for low-lying water hexamer structures,'' {\em The Journal of Physical Chemistry A}, vol.~113, no.~15, pp.~3555--3559, 2009.

\bibitem{ChenGaff}
G.~Chen, T.~Jaffrelot~Inizan, T.~Pl{\'e}, L.~Lagardère, J.-P. Piquemal, and Y.~Maday, ``Advancing force fields parameterization: A directed graph attention networks approach,'' {\em Journal of Chemical Theory and Computation}, vol.~20, no.~13, pp.~5558--5569, 2024.
\newblock PMID: 38875012.

\bibitem{wang2004development}
J.~Wang, R.~M. Wolf, J.~W. Caldwell, P.~A. Kollman, and D.~A. Case, ``Development and testing of a general amber force field,'' {\em Journal of Computational Chemistry}, vol.~25, no.~9, pp.~1157--1174, 2004.

\bibitem{jax2018github}
J.~Bradbury, R.~Frostig, P.~Hawkins, M.~J. Johnson, C.~Leary, D.~Maclaurin, G.~Necula, A.~Paszke, J.~Vander{P}las, S.~Wanderman-{M}ilne, and Q.~Zhang, ``{JAX}: composable transformations of {P}ython+{N}um{P}y programs,'' 2018.

\bibitem{habershon2013ring}
S.~Habershon, D.~E. Manolopoulos, T.~E. Markland, and T.~F. Miller~III, ``Ring-polymer molecular dynamics: Quantum effects in chemical dynamics from classical trajectories in an extended phase space,'' {\em Annual review of physical chemistry}, vol.~64, no.~1, pp.~387--413, 2013.

\bibitem{ple2022routine}
T.~Pl\'e, N.~Mauger, O.~Adjoua, T.~J. Inizan, L.~Lagard\`ere, S.~Huppert, and J.-P. Piquemal, ``Routine molecular dynamics simulations including nuclear quantum effects: From force fields to machine learning potentials,'' {\em Journal of Chemical Theory and Computation}, vol.~19, no.~5, pp.~1432--1445, 2023.

\bibitem{blazhynska2025satellite}
M.~Blazhynska, O.~Adjoua, Z.~Vetter, L.~Lagard{\`e}re, and J.-P. Piquemal, ``Satellite tobacco mosaic virus: Revealing environmental drivers of capsid and nucleocapsid stability using high-resolution simulations,'' {\em bioRxiv}, 2025.
\newblock DOI: 10.1101/2025.01.28.635309.

\bibitem{D1SC00145K}
T.~Jaffrelot~Inizan, F.~Célerse, O.~Adjoua, D.~El~Ahdab, L.-H. Jolly, C.~Liu, P.~Ren, M.~Montes, N.~Lagarde, L.~Lagardère, P.~Monmarché, and J.-P. Piquemal, ``High-resolution mining of the sars-cov-2 main protease conformational space: supercomputer-driven unsupervised adaptive sampling,'' {\em Chem. Sci.}, vol.~12, pp.~4889--4907, 2021.

\bibitem{pytorchpaper}
A.~Paszke, S.~Gross, F.~Massa, A.~Lerer, J.~Bradbury, G.~Chanan, T.~Killeen, Z.~Lin, N.~Gimelshein, L.~Antiga, A.~Desmaison, A.~Kopf, E.~Yang, Z.~DeVito, M.~Raison, A.~Tejani, S.~Chilamkurthy, B.~Steiner, L.~Fang, J.~Bai, and S.~Chintala, ``Pytorch: An imperative style, high-performance deep learning library,'' in {\em Advances in Neural Information Processing Systems 32} (H.~Wallach, H.~Larochelle, A.~Beygelzimer, F.~d\textquotesingle Alch\'{e}-Buc, E.~Fox, and R.~Garnett, eds.), pp.~8024--8035, Curran Associates, Inc., 2019.

\bibitem{constantphreview}
P.~Buslaev, N.~Aho, A.~Jansen, P.~Bauer, B.~Hess, and G.~Groenhof, ``Best practices in constant ph md simulations: Accuracy and sampling,'' {\em Journal of Chemical Theory and Computation}, vol.~18, no.~10, pp.~6134--6147, 2022.
\newblock PMID: 36107791.

\bibitem{Shen2013}
J.~Shen, Y.~Zeng, X.~Zhuang, L.~Sun, X.~Yao, P.~Pimpl, and L.~Jiang, ``Organelle ph in the arabidopsis endomembrane system,'' {\em Molecular Plant}, vol.~6, no.~5, pp.~1419--1437, 2013.

\bibitem{Maria2025}
M.~Maria, S.~Guionnière, N.~Dacquay, C.~Plateau-Holleville, V.~Guillaume, V.~Larroque, J.~Lardé, Y.~Naimi, J.-P. Piquemal, G.~Levieux, N.~Lagarde, S.~Mérillou, and M.~Montes, ``Vtx: Real-time high-performance molecular structure and dynamics visualization software,'' 2025.
\newblock Preprint arXiv:2501.12750.

\end{thebibliography}

\end{document}